\documentclass[nofootinbib]{revtex4}
\usepackage{graphicx}
\usepackage{dcolumn}
\usepackage{amsmath,amssymb,epsfig}
\usepackage{paralist}
\usepackage{comment} 
\usepackage{siunitx}
\usepackage{graphicx}
\usepackage{multirow}
\usepackage{color,soul}
\usepackage{suffix}
\usepackage{mathtools}
\usepackage{tabularx}
\usepackage{tabulary}
\usepackage{array}
\usepackage[percent]{overpic}
\usepackage{placeins}   

\usepackage[breaklinks=True,colorlinks=True, linkcolor=magenta,citecolor=blue,debug=True]{hyperref}

\usepackage{wasysym}

\newcolumntype{C}{>{\centering\arraybackslash}p{0.95cm}}

\renewcommand{\vec}[1]{\boldsymbol{\mathrm{#1}}}

\usepackage{booktabs}

\def\beqn{\begin{eqnarray*}}
\def\eeqn{\end{eqnarray*}}

\newcommand{\be}{\begin{equation}}
\newcommand{\ee}{\end{equation}}
\newcommand{\ba}{\begin{eqnarray}}
\newcommand{\ea}{\end{eqnarray}}

\begin{document}

\title{High-Precision Lunar Corner-Cube Retroreflectors: A Wave‐Optics Perspective}

\author{Slava G. Turyshev}   

\affiliation{ 
Jet Propulsion Laboratory, California Institute of Technology,\\
4800 Oak Grove Drive, Pasadena, CA 91109-0899, USA
}

\begin{abstract}

High-precision corner-cube retroreflectors (CCRs) are critical for advanced lunar laser ranging (LLR) because they enable sub-millimeter-scale measurements of the Earth–Moon distance—a level of precision essential for rigorous tests of relativistic gravitation and for advancing our understanding of lunar geophysics. In this work, we develop a comprehensive two-dimensional Fourier-optics model for single CCRs with apertures ranging from 80–110\,mm. Our model incorporates realistic thermal–mechanical wavefront errors, detailed diffraction effects, and velocity aberration offsets. Our analysis reveals a strong coupling between aperture size and aberration angular  offset: while larger CCRs deliver high on-axis flux under near-ideal conditions, their narrow diffraction lobes suffer significant flux loss at moderate aberration offsets, thereby favoring smaller apertures with broader main lobes. Furthermore, comparisons between solid fused-silica and hollow silicon-carbide (SiC) CCRs show that hollow designs not only achieve competitive or superior photon return—particularly at 1064\,nm, where phase errors are relatively reduced—but also offer nearly an order-of-magnitude mass reduction for the same aperture sizes. These results establish a robust quantitative framework for optimizing CCR designs to perform at the sub-millimeter level under realistic lunar conditions and underscore the advantages of precision hollow SiC CCRs for next-generation LLR operations.

\end{abstract}

\date{\today}
\maketitle


\section{Introduction}

Arrays of corner-cube retroreflectors (CCRs) have been deployed on the Moon since the 1970s, enabling
sub-centimeter Lunar Laser Ranging (LLR) experiments that advance lunar geophysics research
and provide stringent tests of gravitational physics
\citep{Dickey:1994,Murphy_etal_2008,Williams:2009}. Five legacy arrays (Apollo\,11, 14, 15; Lunokhod\,1, 2) each suffer performance degradations
from thermal extremes, dust, and libration-induced tilts, which reduce overall photon return \citep{Dickey:1994,Murphy_etal_2010}. The Apollo reflectors, built from 3.8\,cm fused-silica CCRs (300 prisms on Apollo\,15, 100 each on Apollo\,11 and 14), and Lunokhod’s 14 triangular CCRs (11\,cm edges) jointly exhibit $\sim330\,\mathrm{ps}$ of pulse broadening and
$\pm7\,\mathrm{cm}$ range errors.

By contrast, single large CCRs with 100--200\,mm apertures can reduce  pulse spreading 
\citep{Currie_etal_2011,Otsubo2011,Turyshev-etal:2013,Preston2013}. 
Their smaller collection area, however, requires narrower beam divergence, higher laser power, and more sensitive detectors to reach mm-level precision. To help address these concerns---the Next Generation Lunar Retroreflector (NGLR), based on a single solid CCR with a 100 mm clear aperture made out of a fused silica---was recently developed \citep{Currie_etal_2011} and deployed on the Moon on one of the NASA CLPS landers\footnote{See details at \url{https://www.nasaspaceflight.com/2025/03/blue-ghost-im-2-landings/}} and successfully ranged by several LLR stations.\footnote{See details at \url{https://www.physics.umd.edu/hep/drew/nglr/status_03122025.html}}

Next-generation single CCRs remove libration-driven dispersion by delivering more uniform optical paths, yet velocity aberration, narrower diffraction side lobes, and wavefront distortions continue to complicate the seemingly simple act of ``point and bounce'' \cite{Turyshev-etal:2013,Preston2013,Degnan:2023,Turyshev:CW-LLR:2025}. Although the underlying principle of LLR is straightforward---directing laser pulses toward the Moon and timing their round-trip \cite{Williams:2004,Williams:2009,WilliamsBoggs2020}---the actual optical performance of lunar CCRs is shaped by multiple factors that can compromise the expected return signal.

First, diffraction sets the fundamental beam divergence: a finite aperture unavoidably spreads energy into a central Airy disk and weaker side lobes. In principle, larger apertures have narrower diffraction cones and thus higher on-axis intensity, but any misalignment can cause significant flux loss if the receiver’s field of view is displaced from the main lobe. Second, thermal and mechanical wavefront errors (WFEs) affecting quality of the retro-reflected beam arise from lunar day--night temperature swings of $\sim 390$\,K, which can induce refractive distortions in solid CCRs or mechanical deformations in hollow CCR designs. Such deformations can funnel a portion of the beam into side lobes instead of the central peak, reducing usable return signal. Third, velocity aberration appears because the Moon orbits Earth at $\approx$ 1\,km/s, creating angular offsets on the order of a few microradians between the incoming and reflected beams. Even moderate offsets of $\sim$\,3.9--7.3\,$\mu$rad ($\sim$\,0.8--1.5$''$ \cite{Williams-etal:2023}) can significantly reduce on-axis overlap, especially for large apertures with tight diffraction patterns. Finally, the choice of design itself---solid total-internal-reflection CCRs vs.\ hollow mirror-based trihedrals---affects mass, thermal gradients, mechanical stiffness, and reflectivity. Solid prisms (often fused silica) have historically been used on the Apollo arrays but can experience internal thermal lensing, while hollow CCRs alleviate bulk refraction at the cost of potentially moderate mirror-surface wavefront errors.

Previous flux estimates commonly treated CCR responses via simplistic top-hat or Gaussian beam models, assuming uniform WFEs or no side lobes. In reality, angular offsets and side-lobe contributions can alter the main-lobe intensity in nontrivial ways. Oversimplified approaches may abruptly drop the predicted flux to near zero when the offset exceeds the primary lobe radius, or incorrectly favor large diameters at intermediate offsets. Such extremes omit nuanced physics of how real CCRs distribute light in the far field when subject to partial velocity aberrations and spatially varying wavefront distortions.

We perform a wave‐optics analysis that captures the complex interplay among aperture diameter, WFEs, reflectivity, and velocity aberration in determining the photon return from lunar CCRs. Our approach utilizes a two-dimensional Fourier-propagation algorithm that incorporates realistic WFE maps along with an imposed phase tilt to simulate the relevant velocity aberration angles. By analyzing apertures in the range of 80--110\,mm and comparing both solid fused-silica prisms and hollow mirror-based CCRs, we establish a robust framework for predicting flux behavior under realistic conditions and guiding the optimal design of CCRs for next-generation LLR \cite{Turyshev:CW-LLR:2025}.

This paper is organized as follows. In Section~\ref{sec:geometry}, we describe the geometry and effective apertures of both solid and hollow CCRs, emphasizing their thermal and mechanical properties and outlining our wave‐optics modeling approach. In Section~\ref{sec:velaberr}, we quantify diffraction effects and velocity aberration resulting from lunar orbital motion and Earth's rotation. Section~\ref{sec:thermalWFE} examines the thermal and mechanical wavefront errors induced by extreme lunar temperature swings. In Section~\ref{sec:simulation}, we present our numerical simulation framework—including details on aperture discretization, phase-error modeling, and photon-return scaling—and in Section~\ref{sec:simulation-results} we report the simulated photon-return performance, including comparisons to Apollo LLR measurements. Section~\ref{sec:hollowCCR_Assembly} discusses assembly and qualification procedures for hollow CCRs, such as mirror bonding, alignment tolerances, and environmental testing under differential thermal expansion. In Section~\ref{sec:optimal_ccr_design}, we present optimal CCR instrument design suitable for a robotic deployment on the Moon. The design features two hollow SiC 100 mm CCRs separated by $\sim 0.5$~m on a lander's deck and capable of supporting sub-mm ranging. Finally, in Section~\ref{sec:conclusion} we summarize our findings and conclude.

\section{CCR Geometry, Mechanical and Optical Properties}
\label{sec:geometry}

A CCR returns incident laser radiation exactly antiparallel to its incoming direction using three mutually orthogonal reflective surfaces that converge at a common vertex. Historically, solid CCRs fabricated from high-purity synthetic fused silica have been used in LLR systems \cite{Faller_1970,ArnoldApolloAnalysis,Turyshev:CW-LLR:2025}. More recently, advanced hollow CCR designs that utilize individually mounted planar mirrors in an open trihedral configuration have been developed to address thermal and mechanical limitations of solid prisms \cite{Turyshev-etal:2013,Preston2013,Degnan:2023}.

\subsection{Solid CCRs}

Solid CCRs for lunar applications are constructed from synthetic fused silica (SiO$_2$), chosen for its very low coefficient of thermal expansion ($\alpha_{\text{CTE}} \approx 5.5\times10^{-7}\,\mathrm{K}^{-1}$), high optical homogeneity, and stable refractive indices (with $n=1.4607$ at 532\,nm and $n=1.4496$ at 1064\,nm, see \cite{Malitson1965}). Apollo-era CCR arrays employed prism apertures around 38\,mm, while next-generation systems consider effective apertures on the order of 100\,mm \cite{Currie_etal_2011}. The prism height is typically designed as $\approx$ 70\% of the aperture diameter, ensuring three total internal reflections (TIR) with minimal phase error \cite{Currie_etal_2011,Ciocci2017}. Techniques such as magnetorheological finishing (MRF) and ion-beam figuring (IBF) routinely achieve surface flatness better than $\lambda/10$ (at 632.8\,nm), limiting cumulative wavefront distortion to roughly $\lambda/4$ (at 632.8\,nm) \cite{Preston2013}.

Due to the extreme lunar diurnal thermal cycles—with surface temperatures varying from approximately $390\,\mathrm{K}$ (+117$^\circ$C) during lunar day to $90\,\mathrm{K}$ (–183$^\circ$C) during lunar night—solid CCRs experience significant internal temperature gradients \cite{Murphy_etal_2010}. These gradients induce refractive-index variations on the order of $\Delta n \approx 10^{-5}$ and cause wavefront distortions—primarily defocus and spherical aberrations—that degrade the achievable wavefront quality, thereby lowering the Strehl ratio and reducing the return flux (see Section~\ref{sec:thermalWFE}).

Fresnel reflections at prism surfaces limit the overall effective reflectivity of uncoated solid fused silica CCRs to approximately $\rho \approx 0.92 - 0.93$ \cite{Degnan:2023}, accounting for losses at the three internal reflections and the additional loss from the entry/exit face even with advanced anti-reflection (AR) coatings.  We note TIR itself is essentially lossless; with high-quality AR on the entry/exit face, the \emph{optics-only} two-pass transmission can exceed $\sim$0.98. The $\rho$ values used here include margins for scatter/contamination and are applied consistently to both solid and hollow designs in the link model.

\subsection{Hollow CCRs}

Hollow CCRs consist of three individually mounted, precision-aligned planar mirrors arranged in an open trihedral configuration, thus eliminating refractive-index-induced internal aberrations characteristic of solid designs \cite{Turyshev-etal:2013,Degnan:2023}.  Substrate materials include dimensionally stable, low-expansion glasses or ceramics such as Zerodur ($\alpha_{\text{CTE}}\approx0\pm0.1\times10^{-7}\,\mathrm{K}^{-1}$), fused silica, and silicon carbide (SiC, $\alpha_{\text{CTE}}\approx 2.2\times10^{-6}\,\mathrm{K}^{-1}$).  Among these, SiC offers particularly advantageous mechanical and thermal properties, including a high Young’s modulus, superior thermal conductivity, and moderate thermal expansion (see Table~\ref{tab:ccr-parameters-multi}). Such properties significantly mitigate thermal gradients and associated mechanical distortions during lunar thermal cycling. 

Mirror surfaces employ dielectric multilayer coatings optimized for maximum reflectivity ($\rho \approx 0.98$) at primary operational wavelengths (532\,nm and 1064\,nm) to enhance long-term durability in the harsh lunar environment \cite{Degnan:2023}. To ensure durability under harsh lunar conditions—including micrometeoroid bombardment, solar UV exposure, and solar wind sputtering—mirrors are further protected by metallic coatings of silver (Ag) or gold (Au) beneath dielectric protective overcoats. Silver coatings provide exceptionally high reflectivity ($>98\%$ across visible and near-infrared spectra) but necessitate dielectric overcoats (typically SiO$_2$ or TiO$_2$ layers) to inhibit oxidation and tarnishing. Gold coatings, although marginally lower in reflectivity (95--98\% at infrared wavelengths), offer superior chemical inertness, oxidation resistance, and radiation tolerance, making them ideally suited for prolonged lunar operations.

The open geometry and advanced material selection of hollow CCRs limit internal thermal gradients to approximately 2--5\,$^\circ$C under lunar diurnal cycling \cite{Degnan:2023}. Such gradients predominantly induce lower-order mechanical aberrations—particularly astigmatism and trefoil—resulting in manageable RMS wavefront distortions. Specifications for currently available   products indicate typical RMS wavefront errors of approximately $\lambda/10$ to $\lambda/20$ at 633\,nm, corresponding to roughly 30--60\,nm RMS for apertures around 80\,mm.\footnote{For an example of a hollow CCR, please see Ultra-Stable Hard-Mounted Retroreflector (USHM) by PLX, Inc. at: \\ \url{https://www.plxinc.com/products/improved-ultra-stable-hard-mounted-retroreflector-ushm}} Achieving and maintaining sub-arcsecond mirror alignment tolerances—routinely verified through high-resolution interferometric testing—are critical to maintaining optimal optical performance and ensuring sustained high-accuracy LLR operations \cite{Goodrow2012,Turyshev-etal:2013,Turyshev:CW-LLR:2025}.

\subsection{Quantitative Technical Summary}

Both CCR designs, solid and hollow, require meticulous optical, thermal, and mechanical design analyses to ensure maximal reflection efficiency, minimal wavefront degradation, and long-term reliability under realistic lunar conditions.

\begin{table}[ht!]
\centering
\caption{Combined technical comparison of CCR designs for apertures of 80--100~mm.}
\label{tab:ccr-parameters-multi}
\begin{tabular}{|l|c|c|c|c|}
\hline
\multirow{2}{*}{Parameter} & Solid CCR & \multicolumn{3}{c|}{Hollow CCR} \\
\cline{3-5}
 & Fused silica & Zerodur & Fused silica & SiC \\
\hline\hline
Aperture diameter, $D$ 
  & 80--110~mm 
  & \multicolumn{3}{c|}{80--100~mm} \\
Geometry 
  & Monolithic prism 
  & \multicolumn{3}{c|}{Open three-mirror trihedral} \\
Mirror thickness 
  & Bulk prism ($\sim0.7D$) 
  & 5--8~mm & 5--8~mm & 3--5~mm \\
Reflectivity, $\rho$ 
  & $\approx0.92$ (TIR) 
  & \multicolumn{3}{c|}{$>0.98$ (Protected Ag/Au + dielectric multilayer coatings)} \\
Protective coatings 
  & None required 
  & \multicolumn{3}{c|}{Metallic (Ag/Au) + dielectric multilayers} \\
Thermal gradient, $\Delta T$ 
  & 10--20$^\circ$C 
  & 5--10$^\circ$C & 10--20$^\circ$C & 2--5$^\circ$C \\
Wavefront RMS ($D=80$~mm) 
  & $\lambda/25$--$\lambda/20$ 
  & $\lambda/18$--$\lambda/10$ & $\lambda/20$--$\lambda/10$ & $\lambda/20$--$\lambda/10$ \\
Wavefront RMS ($D=110$~mm) 
  & $\lambda/8$--$\lambda/7$ 
  & $\lambda/9$--$\lambda/7$ & $\lambda/9$--$\lambda/7$ & $\lambda/10$--$\lambda/6$ \\
Dominant aberrations 
  & Defocus, spherical 
  & \multicolumn{3}{c|}{Astigmatism, trefoil} \\
Thermal conductivity 
  & 1.38~W\,m$^{-1}$K$^{-1}$ 
  & 1.46~W\,m$^{-1}$K$^{-1}$ & 1.38~W\,m$^{-1}$K$^{-1}$ & 120--270~W\,m$^{-1}$K$^{-1}$ \\
Coefficient of thermal expansion 
  & $5.5\times10^{-7}$~K$^{-1}$ 
  & $0.05\times10^{-6}$~K$^{-1}$ & $0.55\times10^{-6}$~K$^{-1}$ & $2.2\times10^{-6}$~K$^{-1}$ \\
Young's modulus 
  & 72~GPa 
  & 90~GPa & 72~GPa & 410~GPa \\
Mass (assembled, $D=100$~mm) 
  & 2.0--2.5~kg 
  & 0.6--0.9~kg & 0.5--0.8~kg & 0.4--0.5~kg \\
\hline
\end{tabular}%
\end{table}

Table~\ref{tab:ccr-parameters-multi} presents a quantitative comparison of key technical parameters for solid fused-silica CCRs versus advanced hollow CCR designs fabricated from Zerodur, fused silica, and SiC. While solid CCRs provide stable wavefront performance owing to their monolithic structure, they incur a higher mass (typically 2.0--2.5\,kg for a 100\,mm assembly) and suffer from Fresnel losses. In contrast, hollow CCR assemblies, particularly those using SiC, can achieve up to 80--90\% mass reduction while delivering competitive optical quality and superior thermal performance. The high thermal conductivity and mechanical stiffness of materials such as SiC substantially reduce internal thermal gradients and related aberrations, making hollow CCRs a promising solution for next-generation high-precision LLR \cite{Turyshev:CW-LLR:2025}.

\section{Diffraction and Velocity Aberration}
\label{sec:velaberr}

\subsection{Far-Field Diffraction Pattern of a CCR}

The far-field diffraction pattern produced by a uniformly illuminated circular aperture is described by the Airy diffraction formula \cite{Born-Wolf:1999,Goodman:2017}. Under Fraunhofer diffraction conditions, valid when the observation distance greatly exceeds the Rayleigh range ($z_{\rm Rayleigh}\sim D^2/\lambda$), the normalized intensity distribution at an angular displacement $\theta = \sqrt{\theta_x^2 + \theta_y^2}$, with $\vec\theta = (\theta_x,\theta_y)$ measured from the optical axis, is  given by:
\begin{equation}
I(\theta) = I_0\,\bigg[\frac{2\,J_1\big(\pi D {\theta}/\lambda\big)}{{\pi D \theta}/\lambda}\bigg]^2,
\label{eq:Airy-CCR}
\end{equation}
where $I_0$ denotes the peak on-axis intensity, $D$ is the clear aperture diameter of the CCR, $\lambda$ represents the wavelength of the illuminating laser beam, and $J_1(x)$ is the Bessel function of the first kind, order one. 

\begin{figure}[t!]
\begin{minipage}[b]{.46\linewidth}
\rotatebox{90}{\hskip 30pt  Normalized Intensity}
\includegraphics[width=0.95\linewidth]{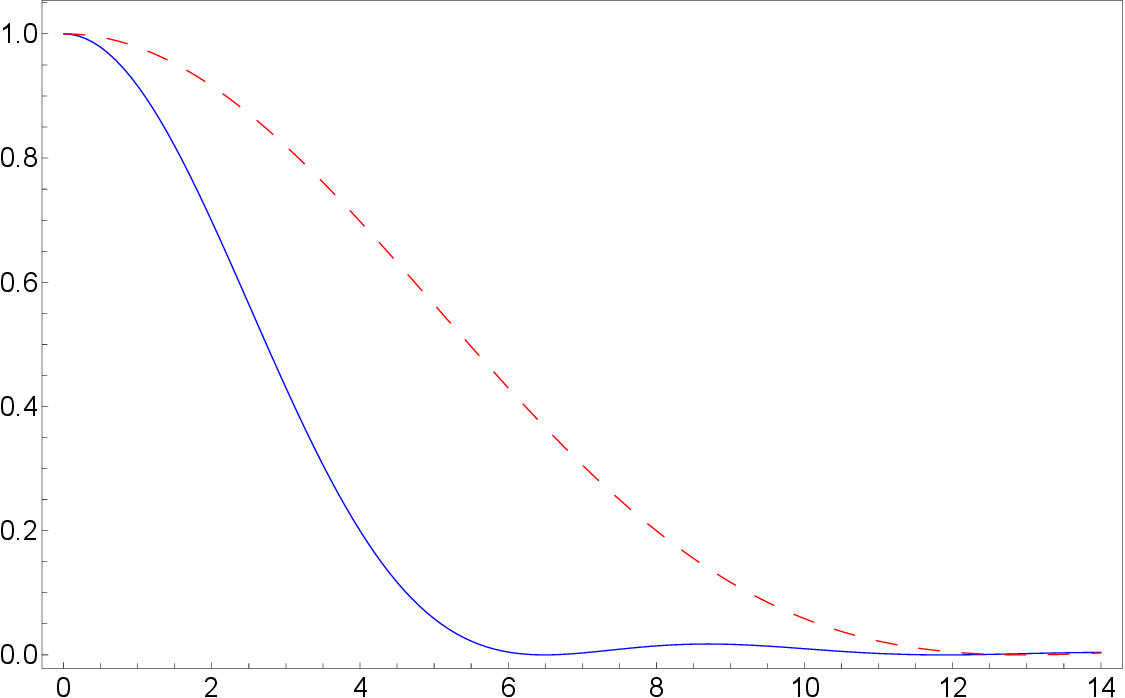}
\rotatebox{0}{\hskip 36pt  Angular displacement $\theta$ ($\mu$rad)}
\end{minipage}
~\,
\begin{minipage}[b]{.46\linewidth}
\rotatebox{90}{\hskip 30pt  Normalized Intensity}
\includegraphics[width=0.95\linewidth]{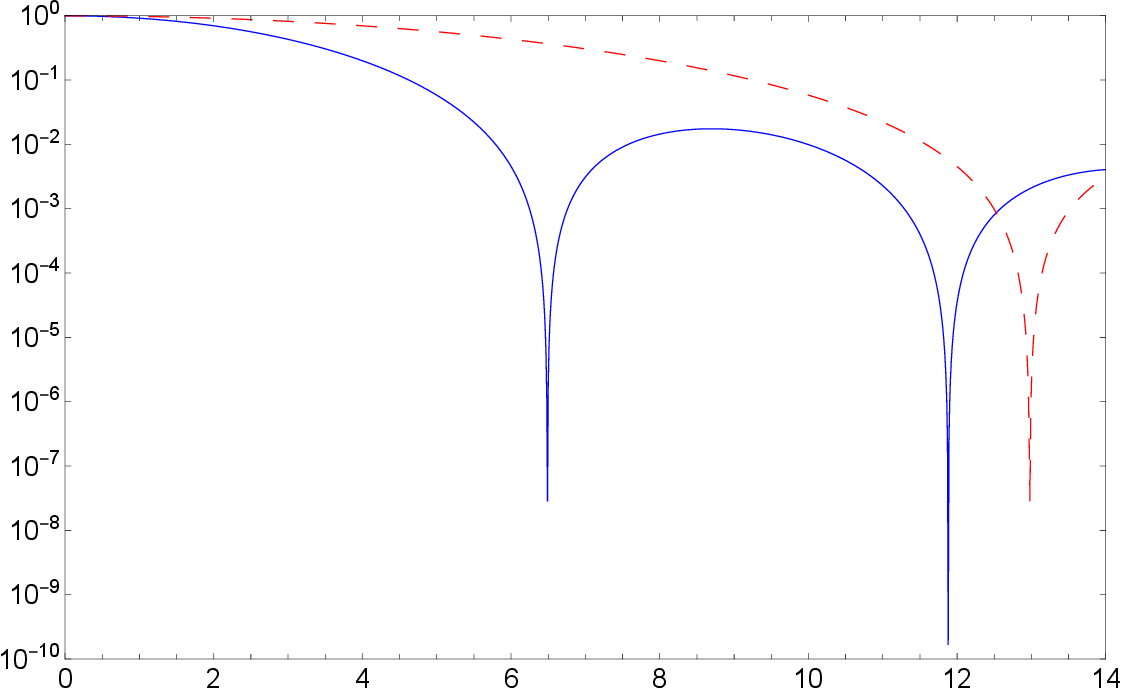}
\rotatebox{0}{\hskip 36pt  Angular displacement $\theta$ ($\mu$rad)}
\end{minipage}
\vskip -2pt
\caption{\label{fig:Airy-Int} The far‐field diffraction pattern of a circular CCR with a 100~mm aperture is plotted on a linear scale (left) and on a logarithmic scale (right). The smooth line corresponds to 532~nm illumination, while the dashed line represents 1064~nm illumination. For 532~nm, the half-power point occurs at \(\theta = 2.737\,\mu\)rad and the first null is observed at \(\theta = 6.485\,\mu\)rad. For 1064~nm, the half-power point is at \(\theta = 5.474\,\mu\)rad and the first null is at \(\theta = 12.969\,\mu\)rad.}
\end{figure}

The first minimum (or dark ring) of the Airy diffraction pattern occurs at the angle:
$\theta_\text{first null} \approx 1.22\,({\lambda}/{D}).$
As an example, at a  $\lambda=532\,\text{nm}$ and $D=100\,\text{mm}$, the first null appears at an angular radius of  $\approx 6.49\,\mu\text{rad}$. At $1064\,\text{nm}$, the same aperture yields a first-null radius at $\approx 12.97\,\mu\text{rad}$, illustrating the linear scaling with wavelength. Figure \ref{fig:Airy-Int} presents the far-field diffraction pattern of a circular CCR with a 100~mm aperture for $\lambda=532$~nm and $\lambda=1064$~nm. 

The Airy pattern is characterized by a central lobe (the Airy disk) and a series of concentric annular side lobes \cite{Born-Wolf:1999}. Quantitatively:
\begin{itemize}
  \item Approximately 83.8\% of the total diffracted power is confined within the central lobe (up to the first null).
  \item The first side-lobe annulus, which extends roughly from \(\theta=1.22\,\lambda/D\) to \(\theta\approx2.23\,\lambda/D\), contains an additional \(\sim 7.2\%\) of the total power.
  \item The remaining \(\sim 9\%\) of the power is distributed among higher-order side lobes.
\end{itemize}
Moreover, the peak intensity of the first side lobe is typically only about 1.75\% of the central peak intensity and diminishes rapidly with increasing order.

This quantitative description of the Airy diffraction pattern, including the half-power and null positions as well as the energy distribution among the central and side lobes, is crucial for understanding and predicting the optical performance of CCRs in high-precision applications.

\subsection{Velocity Aberration in LLR}
\label{sec:vel-aberr}

When a laser pulse is transmitted from an Earth‐based observatory and retroreflected by a lunar CCR, the returning beam is angularly displaced due to the relative motion between the CCR and the station during the round trip. This velocity aberration is a vector phenomenon—it has both magnitude and direction—and it manifests as a tilt in the returning wavefront that shifts the far‐field diffraction pattern relative to the receiver aperture. Such a shift degrades the beam overlap at the receiver and, consequently, the ranging precision.

In an Earth‐centered inertial frame the relative velocity is defined as
\[
\vec{v}_{\mathrm{rel}} = \vec{v}_{\mathrm{CCR}} - \vec{v}_{\mathrm{station}},
\]
and only the component perpendicular to the temporally-varying line‐of‐sight (LOS) direction contributes to the aberration. Let \({\vec n}_{\tt LOS}\) denote the instantaneous unit vector along the LOS; then the effective transverse velocity is
\[
\vec{v}_\perp = \vec{v}_{\mathrm{rel}} - (\vec{v}_{\mathrm{rel}}\cdot{\vec n}_{\tt LOS})\,{\vec n}_{\tt LOS},
\]
with magnitude
$
v_\perp = \|\vec{v}_\perp\|.
$
The round‐trip aberration vector is given by
\begin{equation}
\vec  \alpha \approx 2 \frac{\vec v_\perp}{c} = (\alpha_x,\alpha_y) \qquad\rightarrow\qquad \alpha = \sqrt{\alpha_x^2+\alpha_y^2},
  \label{eq:aberration_rt}
\end{equation}
where $c$ is the speed of light and $(\alpha_x,\alpha_y)$ are the components of the aberration vector.

Accurate LLR requires modeling of velocity aberration effects arising from the relative kinematics between the Earth-based observatory and the lunar CCR. The total angular aberration, $\alpha$, comprises several contributions \cite{Williams-etal:2023}:

\begin{itemize}
    \item \textit{Lunar orbital motion}: The Moon orbits Earth at a mean velocity of  $v_{\mathrm{Moon}}\approx1.023\,\mathrm{km/s}$, producing a one-way angular aberration of $ \alpha_{\mathrm{Moon}} \approx {v_{\mathrm{Moon}}}/{c} \approx 3.41\,\mu\mathrm{rad}$.

    \item \textit{Earth rotation}: Observatories on Earth’s surface rotate at velocities reaching  $\approx 465\,\mathrm{m/s}$ at the equator. The corresponding angular aberration scales as:
$    \alpha_{\mathrm{Earth}}\approx{v_{\mathrm{Earth}}}/{c}\approx1.55\,\mu\mathrm{rad}\,\cos(\text{latitude}).
$

\item \textit{Lunar rotation and librations}: The Moon’s synchronous rotation produces a small linear surface velocity of about $4.62\,\mathrm{m/s}$ at its equator, corresponding to a negligible one-way aberration of $\sim0.015\,\mu\mathrm{rad}$. Lunar librations (up to $\pm7.9^\circ$ longitude, $\pm6.7^\circ$ latitude) and orbital inclination ($5.145^\circ$) introduce slight variations in the effective orientation of lunar CCRs. Although instantaneous aberrations due to librations are typically below $0.1\,\mu\mathrm{rad}$, their cumulative effects over typical observation sessions can reach approximately $0.1$--$0.5\,\mu\mathrm{rad}$.

\end{itemize}

Modern LLR stations may implement a point‐ahead mechanism to pre-compensate for the aberration induced by the station’s motion on the outbound leg. By applying an appropriate point‐ahead angle (ideally equal to the one‐way aberration), the station cancels the outbound contribution, so that the effective aberration observed at the receiver is approximately the one‐way value rather than the full two‐way value.

Table~\ref{tab:velocity-aberration-realistic} summarizes the  aberration contributions for the dominant velocity components. For current LLR stations, under realistic observing conditions (e.g., station latitude, Moon elevation, and libration effects), the observed effective round-trip aberration, is typically in the range of $\alpha \approx 3.9–7.3\, \mu {\rm rad}$ (i.e., $\sim$\,0.8–1.5$''$) \cite{Turyshev-etal:2013,Williams-etal:2023}.

\begin{table}[ht!]
\centering
\caption{Typical velocity aberration contributions for LLR. Contributions from lunar librations are cumulative over a session. }
\label{tab:velocity-aberration-realistic}
\renewcommand{\arraystretch}{1.0}
\begin{tabular}{|l|c|c|c|}
\hline
Aberration Component & Velocity & \multicolumn{2}{c|}{Velocity Aberration} \\\cline{3-4}
                     &    (m/s)      & ($\mu$rad) & (arcsec) \\
\hline\hline
Lunar orbital motion         & $1023$ &  $6.82$   & $1.41$ \\
Earth rotation (equator)     & $465$  &  $3.10$   & $0.64$ \\
Earth rotation ($30^\circ$ latitude) & $402$ &  $2.68$   & $0.55$ \\
Earth rotation ($45^\circ$ latitude) & $328$ &  $2.18$   & $0.45$ \\
Earth rotation ($60^\circ$ latitude) & $233$ &  $1.56$   & $0.32$ \\
Lunar surface rotation       & $4.62$  &  $0.03$   & $0.006$ \\
Lunar librations (cumulative) & --- &  $0.1$--$0.5$  & $0.10$ \\
\hline
Typical aberration, $\alpha$  & --- &   $3.9$--$7.3$~$\mu$rad & $\sim$\,0.8--1.5$''$ \\
\hline
\end{tabular}
\end{table}

Accounting explicitly for these velocity aberration components and their cumulative effect is indispensable for ensuring precise alignment, optimal photon return, and reliable sub-mm-level accuracy in LLR experiments.

\subsection{Implementation in Wave-Optics Models}

In wave-optics models the CCR’s complex field is modified by the phase shift imposed by velocity aberration, which is treated as a vector field. The CCR aperture field is written as
\[
E_{\mathrm{ap}}(x,y) = E_0(x,y)\,\exp\big[i\,\Delta\phi(x,y)\big],
\]
where the phase shift is given by
\[
\Delta\phi(x,y) = k(\vec{\alpha}\cdot\vec{r}) = k\,\big(\alpha_x x + \alpha_y y\big),
\]
with \( k = {2\pi}/{\lambda} \)  and \(\vec{r}=(x,y)\) and $(\alpha_x,\alpha_y)$ are given by (\ref{eq:aberration_rt}). 

This vectorial phase modulation causes the far-field diffraction pattern to shift relative to the receiver’s aperture. Under ideal diffraction-limited conditions (i.e., with a uniform wavefront), the normalized on-axis intensity as a function of angular offset \(\alpha\) is given by the same Airy pattern (\ref{eq:Airy-CCR}), but  $\theta$ with $\alpha$:
\begin{equation}
\frac{I(0,\alpha)}{I_{\mathrm{max}}} = \left[\frac{2\,J_1\left({\pi D \alpha}/{\lambda}\right)}{{\pi D \alpha}/{\lambda}}\right]^2.
\label{eq:abber_Airy}
\end{equation}

It is well known (i.e., \cite{Born-Wolf:1999,Goodman:2017}) that $\approx$ 83.8\% of the total diffracted power from a uniformly illuminated circular aperture is contained within the central Airy disk. An approximate analytical expression for the fraction of power in the main lobe, when the diffraction pattern is laterally shifted by an angular offset \(\alpha\) (with \(\alpha < \theta_1\)), is given by
\begin{equation}
\frac{I(0,\alpha)}{I_{\mathrm{max}}} \approx 0.84\left(\frac{\theta_1 - \alpha}{\theta_1}\right)^2,
\end{equation}
where \(\theta_1 \approx 1.22\,\lambda/D\) is the angular radius of the first diffraction null.  This expression reflects two key aspects: the factor 0.84 represents the fraction of energy originally contained in the central Airy disk, and the quadratic term approximates the reduction in overlap as the beam is laterally displaced by \(\alpha\).

For arbitrary angular offsets—including large values where \(\alpha\) approaches or exceeds the first null—the full Airy function in Eq.\,(\ref{eq:abber_Airy}) for ${I(0,\alpha)}/{I_{\mathrm{max}}}$ must be used to accurately describe the intensity distribution. This rigorous treatment, combined with numerical simulations, captures the behavior of the diffraction pattern in the presence of velocity aberration, including the redistribution of energy into side lobes.

\begin{figure}[t!]
\begin{minipage}[b]{.49\linewidth}
\rotatebox{90}{\hskip 30pt  Normalized Intensity}
\includegraphics[width=0.95\linewidth]{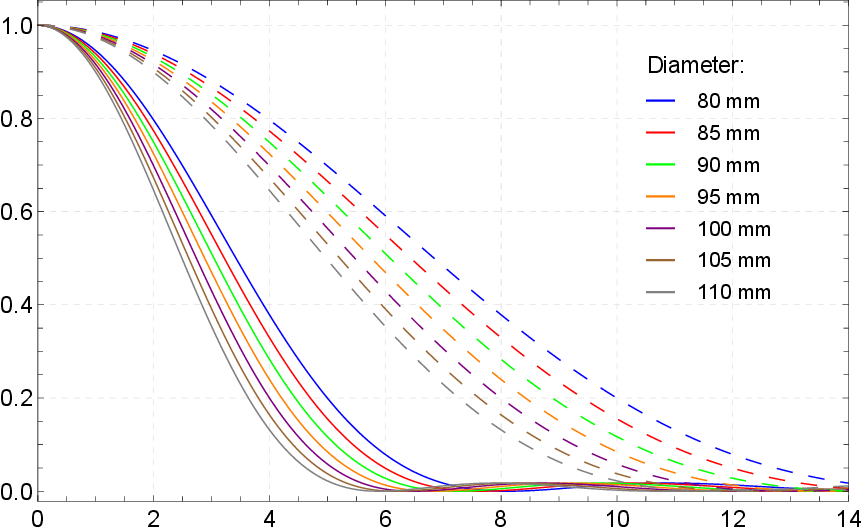}
\rotatebox{0}{\hskip 36pt  Angular displacement $\theta$ ($\mu$rad)}
\end{minipage}
~\,
\begin{minipage}[b]{.49\linewidth}
\rotatebox{90}{\hskip 30pt  Normalized Intensity}
\includegraphics[width=0.95\linewidth]{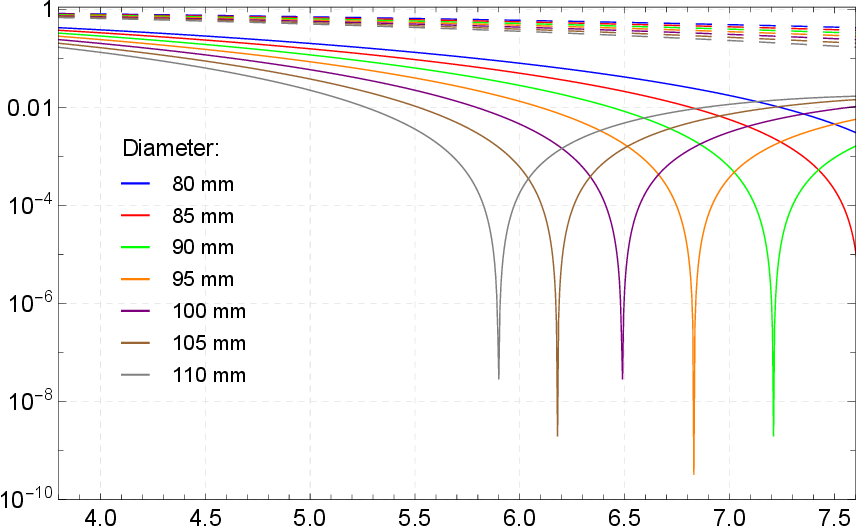}
\rotatebox{0}{\hskip 36pt  Angular displacement $\theta$ ($\mu$rad)}
\end{minipage}
\vskip -2pt
\caption{\label{fig:Airy-zeros-D} The far‐field diffraction pattern of a circular CCR with various apertures as a function of the velocity aberration offset, $\alpha$, as given by (\ref{eq:abber_Airy}). Solid lines correspond to \(\lambda = 532\) nm, while dashed lines are for  \(\lambda = 1064\) nm. From top to bottom, the curves represent CCRs with diameters ranging from 80 mm to 110 mm. Left is linear scale; right is log scale for the angular offset range of $\sim$\,3.8--7.3\,$\mu$rad (i.e., $\sim$\,0.8--1.5$''$) --  region where existing LLR stations are affected by velocity aberration \cite{Williams-etal:2023}. }
\end{figure}

Table~\ref{tab:airy-null-radii} lists the angular radii of the first diffraction null for typical CCR diameters from 80 mm to 110 mm at both wavelengths.  Figure~\ref{fig:Airy-zeros-D} plots the same information highlighting the fact that larger  wavelength, 1064 nm, is less affected by velocity aberration in the LLR-relevant range (see Table~\ref{tab:velocity-aberration-realistic}). 

\begin{table}[h!]
\centering
\caption{Angular radius of first diffraction null, $\theta_1$, for typical CCR diameters $D$  and wavelengths, see Fig.~\ref{fig:Airy-zeros-D}.}
\label{tab:airy-null-radii}
\renewcommand{\arraystretch}{1.0}
\begin{tabular}{|c|c|c|}
\hline
Diameter 
& \multicolumn{2}{c|}{{ $\theta_1$, ($\mu$rad)  }} \\
\cline{2-3}
 (mm) & \(\lambda = 532\) nm & \(\lambda = 1064\) nm\\
\hline\hline
80 & 8.11 & 16.22\\
85 & 7.64 & 15.28\\
90 & 7.21 & 14.42\\
95 & 6.83 & 13.67\\
100 & 6.49 & 12.97\\
105 & 6.18 & 12.36\\
110 & 5.90 & 11.80\\
\hline
\end{tabular}
\end{table}

 In realistic LLR operations, additional wavefront errors (arising from thermal gradients, structural stresses, or fabrication imperfections) redistribute power into the side lobes, so that rigorous numerical simulations (e.g., using a two-dimensional FFT) are required to accurately predict the overlap efficiency.

\subsection{Reduced Impact of Velocity Aberration at 1064 nm}
\label{subsec:vel-aberration-1064}

As we discussed above, the sensitivity of photon return to velocity aberration in nominal LLR operations is significantly reduced at longer wavelengths—particularly at \( \lambda = 1064 \) nm—compared to 532 nm, as seen in  Figure~\ref{fig:Airy-zeros-D}. This reduction is due to the fact that the diffraction-limited full-width half-maximum (FWHM) divergence angle scales with wavelength as
$\theta \approx 1.22\,({\lambda}/{D}),$ so that, for a given CCR aperture diameter \(D\), the diffraction pattern at 1064 nm is wider than that at 532 nm. This broader divergence at 1064 nm means that, for a given angular offset \(\alpha\), a larger fraction of the returned intensity remains within the receiver’s field of view. Table~\ref{tab:velocity-aberration-impact} reports the on‑axis Airy intensity $I(0,\alpha)/I_{\rm max}$ from Eq.~(\ref{eq:abber_Airy}) as a function of \(\alpha\) for both wavelengths for a 100 mm CCR; no receiver‑aperture integration is applied.

\begin{table}[ht!]
\centering
\caption{Fraction of total photon return retained as a function of velocity aberration, \( \alpha \), at 532 nm and 1064 nm, computed from (\ref{eq:abber_Airy}) with $D=100$\,mm. These are \emph{central intensities} of the laterally shifted Airy pattern (no receiver-aperture integration).}
\label{tab:velocity-aberration-impact}
\renewcommand{\arraystretch}{1.1}
\begin{tabular}{|c|c|c|}
\hline
Velocity Aberration 
& \multicolumn{2}{c|}{Retained Flux} \\
\cline{2-3}
  \( \alpha \) (\(\mu\)rad) & \(\lambda = 532\) nm & \(\lambda = 1064\) nm \\
\hline\hline
0  & 100.0\% & 100.0\% \\
2  & 69.8\%  & 91.6\%  \\
4  & 20.0\%  & 69.8\%  \\
6  & 0.45\%  & 43.0\%  \\
8  & 1.45\%  & 20.0\%  \\
\hline
\end{tabular}
\end{table}

One way to mitigate the impact of  velocity aberration at 532\,nm is to intentionally offset one or more dihedral angles in the CCR, thereby introducing a controlled asymmetry into the far-field diffraction pattern  \cite{Turyshev-etal:2013}. Because the Airy disk at 532\,nm is relatively narrow for 80--110\,mm apertures, moderate angular displacements of about 4--7\,\(\mu\)rad can shift the main lobe beyond its first null. Offsetting a dihedral angle by roughly 0.5--1\(''\) re-centers the peak intensity on these typical velocity offsets, but ensuring this offset remains within \(\pm0.2''\) requires sub-arcsecond fabrication tolerances and stable thermal conditions; even minor facet deformations can degrade the intended flux improvement.

Moreover, imposing such a dihedral offset alters the multi-lobed diffraction pattern so that one or two principal lobes carry most of the returning flux, rather than retaining near-rotational symmetry. As a result, a \emph{clocking angle} emerges—the azimuthal orientation of the offset around the CCR’s optical axis—which must be carefully aligned with the average Earth--Moon velocity vector when the CCR is deployed. On uneven lunar terrain, a misalignment of just a few degrees in this rotation can steer the strongest lobes away from Earth-based receivers, negating the benefit of the dihedral offset. Consequently, although dihedral-angle adjustments at 532\,nm can reduce moderate velocity aberrations, they also require more stringent manufacturing tolerances, stricter thermal control, and precise azimuthal alignment to ensure that the re-centered lobes remain within a typical LLR station’s field of view.

By contrast, the broader diffraction pattern at 1064\,nm reduces sensitivity to angular offsets in the 3.9--7.3\,\(\mu\)rad range, making near-infrared wavelengths more robust against small misalignments of the returning beam. Although velocity aberration still shifts the far-field diffraction pattern laterally, the larger Airy disk at 1064\,nm accommodates these displacements with less net flux loss. This difference is key for optimizing CCR designs in long-term LLR applications, particularly where reliable photon return is critical and the constraints of dihedral offsets at shorter wavelengths (e.g., sub-arcsecond machining, strict thermal control) may limit their practicality.

\section{Thermal and Mechanical Wavefront Errors}
\label{sec:thermalWFE}

\subsection{Analytical Formulation of Wavefront Errors}

Wavefront errors (WFEs), denoted as $\phi(x,y)$, quantify deviations in the optical path length across the CCR aperture caused by thermal gradients, mechanical stresses, and fabrication tolerances. These deviations alter the complex aperture field according to
\begin{equation}
E_{\mathrm{aperture}}(x,y) = A_0\,\exp\left[i\,\phi(x,y)\right].
\end{equation}

The overall phase error magnitude is characterized by the root-mean-square (RMS) WFE,
\begin{equation}
\sigma = \frac{\lambda}{2\pi}\sqrt{\left\langle\left(\phi(x,y)-\bar{\phi}\right)^2\right\rangle_{\mathrm{aperture}}},
\end{equation}
where $\bar{\phi}$ is the mean phase over the aperture.

For $\sigma \ll \lambda$, the Mar\'echal approximation quantifies the degradation in optical performance via the Strehl ratio \cite{Born-Wolf:1999,Goodman:2017}:
\begin{equation}
S = \exp\Big[-\Big(\frac{2\pi\sigma}{\lambda}\Big)^2\Big],
\label{eq:strehl}
\end{equation}
with $\sigma$ in meters and $\lambda$ the operational wavelength.

It is important to note that (\ref{eq:strehl}) assumes that phase errors across the CCR aperture are completely random and uncorrelated—that is, they have zero spatial correlation length. Under this idealized condition, the Marechal approximation accurately predicts an exponential reduction in on-axis intensity.

In practical CCRs, however, phase errors arising from surface polishing, bonding, and thermal gradients often exhibit significant spatial correlations. Rather than uniformly attenuating the main lobe, these correlated errors tend to redistribute a portion of the optical power into distinct ``error lobes'' whose structure deviates from the conventional Airy sidelobes. Consequently, the effective Strehl ratio measured in a fabricated CCR may differ from the ideal prediction of (\ref{eq:strehl}). Quantitative characterization of such behavior typically requires a detailed statistical analysis—commonly implemented via a Zernike polynomial decomposition—to capture both the amplitude and the spatial correlation properties of the induced phase errors.

Furthermore, because the overall wavefront distortion in CCRs generally comprises multiple Zernike modes beyond simple defocus and spherical aberration, higher-order aberrations (e.g., astigmatism and trefoil) must be modeled numerically to accurately predict optical performance under realistic lunar environmental conditions.

\subsection{Wavefront Error Modeling}

Wavefront error maps were generated for CCRs with diameters ranging from 80 mm to 110 mm in 5 mm increments. These simulations employ a Zernike polynomial decomposition with mode weightings that capture physically realistic thermal, mechanical, and optical distortions. The model incorporates both refractive index variations and mechanically induced deformations to characterize wavefront errors under lunar conditions.

\subsubsection{Solid CCRs (Fused Silica)}

Wavefront distortions in solid fused-silica CCRs arise from internal thermal gradients due to lunar diurnal temperature fluctuations \cite{Goodrow2012,Murphy:2013,Murphy2013Polarization}. Although these monolithic prisms provide high thermal and mechanical stability, they are susceptible to refractive distortions caused by internal temperature gradients. Over a full lunar day-night cycle, the temperature can vary by \(\Delta T \approx 290\) K, leading to refractive index perturbations of \(\Delta n \sim 10^{-5}\). Such variations primarily introduce optical path distortions in the form of defocus (\(Z_2^0\)) and spherical aberration (\(Z_4^0\)), degrading the diffraction-limited retroreflection.

Fused silica’s low coefficient of thermal expansion (CTE \(\approx 5.5 \times 10^{-7}\,\mathrm{K}^{-1}\)) and moderate thermo-optic coefficient (\(\sim 10^{-5}\,\mathrm{K}^{-1}\)) help reduce bulk deformations but do not eliminate thermally induced optical aberrations. As shown in Table~\ref{tab:wfe_ccrs}, the resulting RMS WFEs for solid CCRs increase with aperture size—ranging from $\approx$ \,30--50 nm for an 80\,mm CCR to 60--80 nm for a 110\,mm CCR at 532\,nm. At 1064\,nm these errors correspond to roughly \(\lambda/35\) to \(\lambda/21\) for the 80\,mm aperture and \(\lambda/18\) to \(\lambda/13\) for the 110\,mm aperture, leading to improved Strehl ratios and higher photon return efficiency. Consequently, solid CCRs remain particularly well-suited for LLR at near-IR wavelengths.

\subsubsection{Hollow CCRs (Silicon Carbide, SiC)}

Hollow CCRs do not experience refractive distortions but are more susceptible to mechanically induced WFEs. Although, hollow CCR designs eliminate bulk refractive distortions but are susceptible to misalignment due to mechanical flexure, mounting stress, and differential thermal expansion effects. Unlike solid CCRs, which experience uniform refractive perturbations, hollow CCRs exhibit higher-order optical aberrations, primarily astigmatism (\(Z_2^{\pm2}\)) and trefoil (\(Z_3^{\pm3}\)), due to anisotropic thermal expansion and structural flexure. The magnitude of these distortions depends on the mechanical stability of the mirror mounts and alignment tolerances.

Although hollow CCRs typically undergo smaller temperature differentials ($\Delta T \approx 5$--$10^\circ\mathrm{C}$), they rely on SiC substrates with a notably higher coefficient of thermal expansion ($\alpha \approx 2 \times 10^{-6}\,\mathrm{K}^{-1}$) than fused silica. This difference in thermal expansion leads to mechanically induced deformations that can significantly increase the RMS WFE. As indicated in Table~\ref{tab:wfe_ccrs}, the RMS WFE at 532\,nm for hollow CCRs ranges from approximately 20--35 nm for an 80\,mm aperture and increases to about 35--50 nm for a 110\,mm aperture. Since the same physical distortions correspond to fewer waves at 1064\,nm, the relative phase error is effectively reduced at longer wavelengths, mitigating wavefront distortions and improving overall optical performance in the near‐infrared.

Rigid mounting solutions with sub-arcsecond alignment tolerances, such as those implemented in PLX’s Ultra-Stable Hard-Mounted (USHM) design\footnote{For  more information on design and properties of PLX’s Ultra-Stable Hard-Mounted (USHM) CCRs, see \url{www.plxinc.com/products/improved-ultra-stable-hard-mounted-retroreflector-ushm}}, mitigate some of these mechanical instabilities. Under controlled alignment conditions, hollow CCRs can achieve optical performance comparable to solid CCRs, particularly in applications where mass constraints drive the need for lightweight mirror-based CCRs.

\subsubsection{Technical Summary}

Table~\ref{tab:wfe_ccrs} summarizes the estimated RMS WFEs for both solid and hollow CCRs at 532 nm and 1064 nm. Table~\ref{tab:wfe-strehl-realistic} presents the median achievable WFE values the corresponding Strehl ratios (\ref{eq:strehl}), which quantify the impact of wavefront distortions on the diffraction-limited performance of solid and hollow CCRs.

High-quality fused silica CCRs typically achieve a transmitted WFE ranging from \(\lambda/10\) to \(\lambda/20\) at 633\,nm ($\approx$\,63\,nm to 31.6\,nm RMS), and CCRs with sub-arcsec beam deviation across the full aperture are now commercially available.\footnote{See Precision Optical Inc., \url{https://www.precisionoptical.com/precision-optics/custom-prism/corner-cube-retroreflector/}}

As for hollow SiC CCRs, for instance, USHM designs from PLX Inc. have recenly demonstrated a reflected WFE as low as \(\lambda/10\) peak-to-valley (P–V) at 633\,nm. Converting this to RMS—given that the RMS is typically $\sim$ 1/5 to 1/6 of the P-V value—yields an RMS WFE of $\approx$\,10–13\,nm for hollow CCRs, further improving the values in Table~\ref{tab:wfe_ccrs}.

\begin{table}[h]
\centering
\caption{Typical WFEs for solid fused silica and hollow SiC CCRs. WFE ranges are in nanometers, with approximate wave counts at 532\,nm and 1064\,nm.}
\label{tab:wfe_ccrs}
\renewcommand{\arraystretch}{1.15}
\begin{tabular}{|c|ccc|ccc|}
\hline
\multirow{2}{*}{Diameter (mm)} 
& \multicolumn{3}{c|}{Solid CCR (Fused Silica)} 
& \multicolumn{3}{c|}{Hollow CCR (SiC)} \\
\cline{2-7}
 & WFE (nm) & $\lambda=532$\,nm & $\lambda=1064$\,nm
 & WFE (nm) & $\lambda=532$\,nm & $\lambda=1064$\,nm \\
\hline\hline

80  & 30--50          
    & $\lambda/18$--$\lambda/11$
    & $\lambda/35$--$\lambda/21$

    & 20--35 
    & $\lambda/27$--$\lambda/15$
    & $\lambda/53$--$\lambda/30$ \\

85  & 35--55          
    & $\lambda/15$--$\lambda/10$ 
    & $\lambda/30$--$\lambda/19$

    & 22--38
    & $\lambda/24$--$\lambda/14$
    & $\lambda/48$--$\lambda/28$ \\

90  & 40--60          
    & $\lambda/13$--$\lambda/9$ 
    & $\lambda/27$--$\lambda/18$

    & 25--40
    & $\lambda/21$--$\lambda/13$
    & $\lambda/42$--$\lambda/27$ \\

95  & 45--65          
    & $\lambda/12$--$\lambda/8$ 
    & $\lambda/24$--$\lambda/16$

    & 28--42
    & $\lambda/19$--$\lambda/12$
    & $\lambda/38$--$\lambda/25$ \\

100 & 50--70          
    & $\lambda/11$--$\lambda/8$ 
    & $\lambda/21$--$\lambda/15$

    & 30--45
    & $\lambda/18$--$\lambda/12$
    & $\lambda/36$--$\lambda/24$ \\

105 & 55--75          
    & $\lambda/10$--$\lambda/7$  
    & $\lambda/19$--$\lambda/14$

    & 32--48
    & $\lambda/17$--$\lambda/11$
    & $\lambda/34$--$\lambda/23$ \\

110 & 60--80          
    & $\lambda/9$--$\lambda/7$  
    & $\lambda/18$--$\lambda/13$

    & 35--50
    & $\lambda/15$--$\lambda/11$
    & $\lambda/30$--$\lambda/22$ \\

\hline
\end{tabular}
\end{table}

\begin{table}[ht]
\centering
\caption{Representative RMS WFEs (taken as median values from Table~\ref{tab:wfe_ccrs}) and corresponding Strehl ratios for solid and hollow CCRs at 532\,nm and 1064\,nm.}
\label{tab:wfe-strehl-realistic}
\renewcommand{\arraystretch}{1.0}
\begin{tabular}{|c|c|cc|c|cc|}
\hline
\multirow{3}{*}{Diameter (mm)} 
& \multicolumn{3}{c|}{Solid CCR} 
& \multicolumn{3}{c|}{Hollow CCR} \\
\cline{2-7}
& WFE & \multicolumn{2}{c|}{Strehl ratio} 
& WFE & \multicolumn{2}{c|}{Strehl ratio} \\
\cline{3-4}\cline{6-7}
& (nm) & 532 nm & 1064 nm 
& (nm) & 532 nm & 1064 nm \\
\hline\hline
80  & 40  & 0.80 & 0.95  & 28  & 0.90 & 0.97 \\
85  & 45  & 0.75 & 0.93  & 30  & 0.88 & 0.97 \\
90  & 50  & 0.71 & 0.92  & 33  & 0.86 & 0.96 \\
95  & 55  & 0.66 & 0.90  & 35  & 0.84 & 0.96 \\
100 & 60  & 0.61 & 0.88  & 38  & 0.82 & 0.95 \\
105 & 65  & 0.55 & 0.86  & 40  & 0.80 & 0.95 \\
110 & 70  & 0.51 & 0.84  & 43  & 0.77 & 0.94 \\
\hline
\end{tabular}
\end{table}

The improvement at 1064\,nm arises from the inverse relationship between optical path errors and wavelength. Since a given RMS WFE introduces fewer optical phase cycles at 1064\,nm than at 532\,nm, the relative impact of wavefront errors is reduced. This property makes hollow CCRs a competitive alternative for near-IR applications, particularly in scenarios where mass efficiency and thermal stability are primary concerns.

The WFE of CCRs deployed on the lunar surface is strongly influenced by diurnal thermal gradients, mounting stress, and dust exposure. For example, 100\,mm solid fused silica CCRs typically exhibit RMS WFEs of $\sim$\,50--70\,nm under Earth conditions, whereas under expected lunar conditions these errors can increase to $\approx$\,50--100\,nm due to enhanced thermal and mechanical stresses. In contrast, 100\,mm hollow CCRs based on SiC typically achieve WFEs of 30--45\,nm under Earth conditions, and although some additional distortion might occur on the Moon, their superior thermal stability and lightweight design help maintain WFEs in nearly the same range. Advanced hollow SiC designs thus show promise for maintaining $\lambda/10$ RMS or better performance even under extreme lunar daytime conditions. Ongoing advancements in SiC mirror fabrication, interferometric alignment, and mechanical mounting techniques could further reduce mechanical distortions in hollow CCRs, enhancing their suitability for high-precision LLR.

\subsection{CCR Mass Model}

Accurately estimating CCR mass is critical for LLR, particularly when multiple units or entire arrays are deployed. Modern solid fused-silica CCRs—including the prism and its supporting structure—typically follow a cubic mass scaling, reflecting volumetric growth with diameter. A representative relation is
\begin{equation}
m_{\text{solid}} \approx 2.0\,\text{kg} \Big(\frac{D}{100\,\text{mm}}\Big)^3,
\label{eq:mass_solid}
\end{equation}
consistent with recent next-generation CCR developments for robotic missions (see, e.g., \cite{Currie_etal_2011,Turyshev-etal:2013}). In Eq.~(\ref{eq:mass_solid}), $D$ is the overall CCR diameter, and 2.0\,kg refers to a notional 100\,mm-diameter fused-silica assembly (including its mount).

Table~\ref{tab:ccr_masses} summarizes the estimated masses for typical CCR diameters.

\begin{table}[ht!]
\centering
\caption{Estimated CCR masses (optics only) for various diameters (solid: fused silica; hollow: silicon carbide.)}
\label{tab:ccr_masses}
\begin{tabular}{|l|ccccccc|}
\hline
Diameter (mm)           & 80   & 85   & 90   & 95   & 100  & 105  & 110  \\
\hline\hline
Solid CCR mass (kg)     & 1.02 & 1.23 & 1.46 & 1.71 & 2.00 & 2.32 & 2.66 \\
Hollow CCR mass (kg)    & 0.12 & 0.13 & 0.15 & 0.16 & 0.18 & 0.20 & 0.22  \\
\hline
\end{tabular}
\end{table}

By contrast, hollow CCRs constructed from SiC gain their mass advantage from thin ($\approx$\,2\,mm) reflective mirrors, supported by minimal structural elements. This yields a more area-driven (rather than volume-driven) mass relation:
\begin{equation}
m_{\text{hollow}} \approx 0.18\,\text{kg} \Big(\frac{D}{100\,\text{mm}}\Big)^2,
\label{eq:mass_hollow}
\end{equation}
reflecting precision mirror-fabrication methods \cite{Degnan:2023}. Compared to comparable-size solid CCRs, hollow SiC CCRs can achieve mass reductions approaching an order of magnitude---an important factor in lunar mission payload constraints (optics-only; assembly mass including frame/hood is 0.4–0.5\,kg at $D=100$\,mm).

These estimates, as shown in Table~\ref{tab:ccr_masses}, are broadly consistent with both current LLR hardware and design studies aimed at deploying smaller CCRs on the Moon. Further details on mechanical margins, dimensional tolerances, and robust optical mounting can be found in \cite{Murphy:2013,Turyshev-etal:2013}, where low thermal expansion and mechanical stiffness are highlighted as key to maintaining precise optical alignment under lunar day--night temperature swings (with $\Delta T\approx 290\,^\circ\mathrm{C}$).

\subsection{Lunar dust: risk and mitigation}
\label{sec:dust}
Lunar fines transported ballistically or electrostatically can \emph{occlude} aperture area and add diffuse scatter. In our link model (e.g., (\ref{eq:photon-return-alpha2})), an areal-occlusion fraction $f$ scales the clear aperture as $D\rightarrow D\sqrt{1-f}$ so that the on-axis flux scales approximately as $F\propto (1-f)^2$ at fixed $\lambda$ and $\alpha$. Thus $f=0.10$ reduces ideal return by $\sim19\%$, before any additional scatter is considered. For \emph{solid} CCRs, the dominant dust effect is added loss/scatter at the entry/exit face; for \emph{hollow} CCRs, dust can settle on exposed mirrors unless mitigated. We therefore adopt a low-profile baffle lip around the aperture, a $5$--$10^\circ$ downward cant of the cavity relative to local horizontal, conductive/ESD-friendly finishes, and—where available—an electrodynamic dust shield (EDS) on the aperture frame.\footnote{For a recent lander‑scale EDS implementation, see Firefly’s Blue Ghost Mission 1: \url{https://fireflyspace.com/missions/blue-ghost-mission-1/}.} A simple deployable cover further limits passive deposition. While detailed dust transport is out of scope, the parametric $f$-model enables site-specific updates in the link budget.

\section{Wave-Optics Performance Simulation}
\label{sec:simulation}

Here we develop a full-wave numerical simulation to model the optical response of lunar CCRs---incorporating diffraction, WFEs, and velocity aberration---\emph{following} established considerations for lunar CCR performance~\cite{Degnan:2023} and standard wave/Fourier optics~\cite{Born-Wolf:1999,Goodman:2017}. The model provides a quantitative framework for evaluating photon return efficiency under operational conditions, accounting for aperture-dependent optical distortions, diffraction-induced beam divergence, velocity aberration, and side-lobe energy redistribution effects.

\subsection{Computational Model and Numerical Implementation}
\label{sec:sim-set-up}

A structured computational framework is employed to model the photon return characteristics of lunar CCRs, incorporating diffraction-limited beam propagation, wavefront distortion effects, and velocity aberration-induced misalignment. The methodology consists of several key components:

\subsubsection{Aperture Discretization and Grid Resolution}

Each CCR is modeled as a circular aperture of diameter \(D\), sampled onto a uniform Cartesian grid:
\begin{equation}
    x, y \in \big[-\textstyle{\frac{1}{2}}D, +\textstyle{\frac{1}{2}}D\big].
\end{equation}
The aperture field is discretized using a grid resolution of \(N \times N\) points, where \(N \geq 512\) (typically 1024), ensuring accurate representation of both the aperture function and WFE distribution. The complex aperture field is initialized as:
\begin{equation}
    E_{\text{ap}}(x,y) =
    \begin{cases}
        E_0, & r \leq \textstyle{\frac{1}{2}}D, \\
        0,   & r > \textstyle{\frac{1}{2}}D.
    \end{cases}
\end{equation}
Grid resolution is selected to prevent aliasing artifacts in the Fourier transform used for far-field computation, ensuring that diffraction effects and wavefront distortions are accurately resolved.

\subsubsection{Wavefront Error Modeling and Phase Distortion}

WFEs are introduced as a spatially varying phase distortion \(\phi(x,y)\), which represents optical path deviations relative to an ideal retroreflecting wavefront. These distortions are decomposed using a Zernike polynomial expansion:
\begin{equation}
    \phi(x,y) = \sum_{n,m} a_{n,m} Z_{n,m}(x,y),
\end{equation}
where the coefficients \(a_{n,m}\) are determined to match empirical RMS wavefront errors (\(\sigma\)), as detailed in Table~\ref{tab:wfe-strehl-realistic}. The resulting distorted aperture field is expressed as:
\begin{equation}
    E_{\text{ap}}(x,y) = E_0 \exp\left[i\,\phi(x,y)\right], \quad \text{for} \quad r \leq \textstyle{\frac{1}{2}}D.
\end{equation}

Wavefront error characteristics vary significantly between solid and hollow CCRs:

\begin{itemize}
    \item \textit{Solid CCRs}: Lunar temperature fluctuations induce refractive index variations within fused silica, introducing defocus and spherical aberration, which degrade retroreflection and redistribute power into diffraction side lobes.
    \item \textit{Hollow CCRs}: The absence of an internal optical path eliminates thermal lensing, but mechanical stress, mirror misalignment, and flexure introduce astigmatism and trefoil distortions, affecting beam collimation efficiency.
\end{itemize}

In this work the Zernike coefficients $a_{n,m}$ are \emph{assigned parametrically}—not fitted to a specific interferogram—to reproduce the RMS values in Tables~\ref{tab:wfe_ccrs}--\ref{tab:wfe-strehl-realistic} while emphasizing mode content typical of each design. For solid fused-silica CCRs, we weight lower-order defocus and spherical aberration ($Z_2^0$, $Z_4^0$), whereas for hollow SiC CCRs we emphasize astigmatism and trefoil ($Z_2^{\pm 2}$, $Z_3^{\pm 3}$), consistent with thermo-mechanical behavior discussed in Sec.~\ref{sec:thermalWFE}. The full coefficient sets used to generate Figs.~\ref{fig:real-CCR-532nm}--\ref{fig:real-CCR-1064nm} are available from the author for reproducibility.

\subsubsection{Velocity Aberration and Beam Misalignment}

Velocity aberration arises from the relative motion between the Earth-based laser station and the lunar CCR array. The Moon's orbital velocity (\(\sim1\) km/s), Earth's rotation, and lunar librations induce an angular offset \(\alpha\) between the transmitted and returned beam directions, see Table~\ref{tab:velocity-aberration-realistic}. We implement the round-trip velocity aberration as a lateral phase tilt, so that
 this offset is modeled as  
\begin{equation}
    E_{\text{ap}}(x,y) \rightarrow E_{\text{ap}}(x,y) \exp\big[i k (\vec x \cdot \vec  \alpha)\big],
\end{equation}
where \(k = 2\pi/\lambda\) is the optical wave number. The resulting lateral displacement shifts the far-field diffraction pattern by the corresponding angular displacement \(\vec \alpha\), simulating the effects of velocity aberration.

\subsubsection{Far-Field Diffraction Computation}

The aperture field is propagated into the far-field domain using the Fraunhofer diffraction integral, computed via a two-dimensional fast Fourier transform (FFT):
\begin{equation}
    E_{\text{far}}(k_x, k_y) = \mathcal{F}\{E_{\text{ap}}(x,y)\}.
\end{equation}
The far-field intensity distribution is given by:
\begin{equation}
    I_{\text{far}}(k_x, k_y) = \big|E_{\text{far}}(k_x, k_y)\big|^2.
\end{equation}
Here, \(k_x\) and \(k_y\) represent angular spatial frequencies that correspond to the far-field propagation angles. The computational domain is selected to include all significant diffraction features, ensuring that the primary return beam and diffraction side lobes are captured.

\subsubsection{Scaling of Photon Return Efficiency}
\label{subsec:photon-return-scaling}

When the returned beam from a CCR is perfectly aligned (\(\alpha=0\)), the total photon flux \(F(D)\) collected by an Earth-based telescope can be approximated by
{}
\begin{equation}
    F(D) 
    \;\propto\; 
    \frac{D^4}{\lambda^2}
    \; 
    \rho
    \;
    \frac{I_{\mathrm{far}}(0,0)}{I_{\mathrm{far,ideal}}(0,0)},
    \label{eq:photon-return-scaling}
\end{equation}
where:
 \(D^2\) represents the \textit{geometric collecting area} of the CCR; \(\bigl({D}/{\lambda}\bigr)^2\) captures  \textit{diffraction-limited beam collimation} (a larger \(D\) narrows the far-field divergence); \(\rho\) is the \textit{reflectivity} (e.g.\ 
          \(\rho \approx 0.92\) for solid fused-silica or 
          \(\rho \approx 0.95\) for hollow mirror-based CCRs);  \({I_{\mathrm{far}}(0,0)}/{I_{\mathrm{far,ideal}}(0,0)}\equiv S\) is the  \textit{Strehl ratio} (\ref{eq:strehl}), reflecting wavefront errors, thermal distortions,  side-lobe power, and other non-idealities (assuming \(\alpha=0\)).

Eq.~\eqref{eq:photon-return-scaling} assumes that all return power remains within the main Airy lobe. However, there are additional losses in LLR due to beam divergence, side-lobe redistribution, and velocity aberration misalignment. In practice, however, \textit{velocity aberration} caused by the Earth's rotation and the Moon's orbital motion introduces a small angular offset \(\alpha\). Once the retroreflected beam is \emph{tilted} by \(\alpha\), the on-axis intensity at the telescope can drop sharply if \(\alpha\) exceeds the approximate central-lobe radius 
(\(\sim 1.22 \,{\lambda}/{D}\)). To quantify this reduction, we define an  \textit{overlap efficiency} \(\eta(\alpha)\):\begin{equation}
    \eta(\alpha) 
    \;=\; 
    \frac{I_{\mathrm{far}}(0,0;\alpha)}%
         {I_{\mathrm{far,ideal}}(0,0)}=\left[\frac{2\,J_1\left({\pi D \alpha}/{\lambda}\right)}{{\pi D \alpha}/{\lambda}}\right]^2,
    \label{eq:eta-alpha-def}
\end{equation}
where \(I_{\mathrm{far}}(0,0;\alpha)\) is the measured on-axis intensity when the beam 
is tilted by \(\alpha\).  Thus, we can write the more general formula for the flux in the case of  a \(\textit{nonzero}\)  \(\alpha\) from (\ref{eq:photon-return-scaling}):
{}
$    F(D,\alpha)
    \;\propto\;
   ({D^4}/{\lambda^2})\, \rho \,  S \,
    \eta(\alpha)$, or
{}
\begin{equation}
    F(D,\alpha)
    \;\propto\;
    \frac{D^4}{\lambda^2}
    \;
    \rho
    \;
    S(\lambda)
    \;
  \left[\frac{2\,J_1\left({\pi D \alpha}/{\lambda}\right)}{{\pi D \alpha}/{\lambda}}\right]^2.
    \label{eq:photon-return-alpha}
\end{equation}

Hence, while enlarging \(D\) improves beam collimation (\(D/\lambda\) grows), it also increases the risk that even moderate \(\alpha\) can nudge the main lobe outside the 
detector field.  In this sense, \({S}\) (\textit{Strehl ratio}) captures wavefront-limited performance at \(\alpha=0\), while \({\eta(\alpha)}\) (\textit{overlap efficiency}) encodes additional flux losses from velocity‐aberration  misalignment.

\subsubsection{Flux Normalization and Comparative Analysis}

To enable direct performance comparison across different CCR configurations, photon return flux values are normalized relative to the maximum observed flux:
$
    F_{\text{norm}}(D,\alpha) = {F(D,\alpha)}/{\max_{D,\text{design}}[F(D,\alpha)]}
    $, yielding
{}
\begin{equation}
      F_{\text{norm}}(D,\alpha)
    \;\propto\;
    \Big(\frac{D}{D_{\tt ref}}\Big)^4  \; \rho \; S(\lambda) \;
  \left[\frac{2\,J_1\left({\pi D \alpha}/{\lambda}\right)}{{\pi D \alpha}/{\lambda}}\right]^2.
    \label{eq:photon-return-alpha2}
\end{equation}
This normalization facilitates direct evaluation of CCR performance across different aperture sizes, reflectivity levels, wavefront qualities, and velocity aberration conditions.

\section{CCR Performance Simulation}
\label{sec:simulation-results}

A numerical simulation was conducted to explore the performance of CCRs across a range of velocity aberrations (\(\alpha \in \{0,1,...,8\}\) \(\mu\)rad) and diameters spanning \(D \in \{80, 85, 90, 95, 100, 105, 110\}\) mm. The results characterize photon return efficiency as a function of aperture size, wavefront distortion, and lunar motion effects, serving as a benchmark for evaluating next-generation CCR designs for high-precision LLR applications.

\subsection{Ideal CCRs: No Wavefront Errors, Perfect Reflectance}

\begin{figure}[t!]
\begin{minipage}[b]{.48\linewidth}
\rotatebox{90}{\hskip 30pt  Normalized Intensity}
\includegraphics[width=0.95\linewidth]{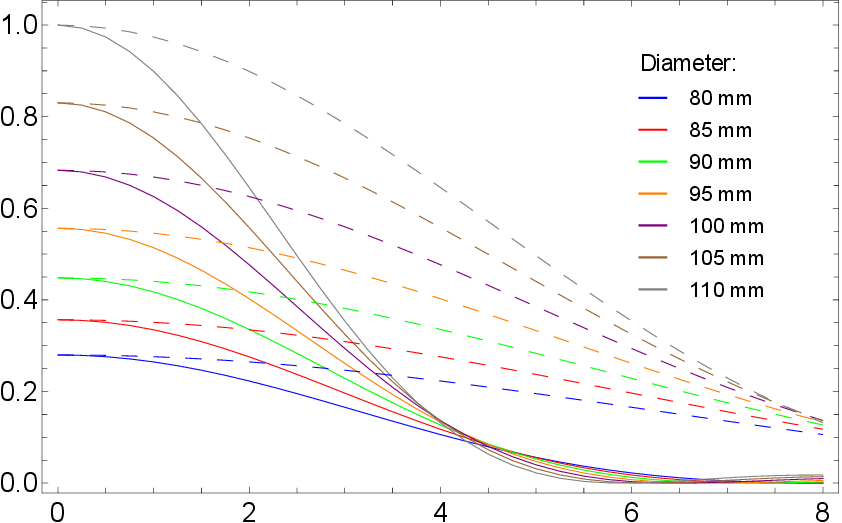}
\rotatebox{0}{\hskip 36pt  Angular displacement $\theta$ ($\mu$rad)}
\end{minipage}
~\,
\begin{minipage}[b]{.50\linewidth}
\rotatebox{90}{\hskip 30pt  Normalized Intensity}
\includegraphics[width=0.95\linewidth]{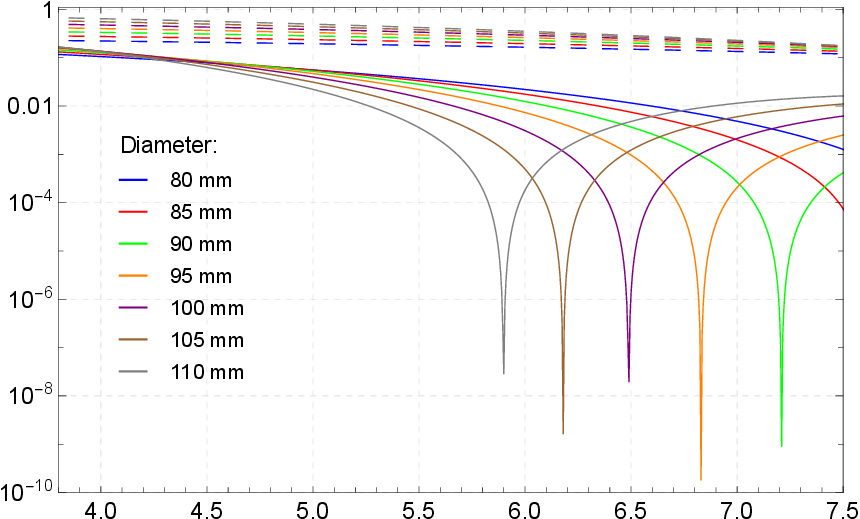}
\rotatebox{0}{\hskip 36pt  Angular displacement $\theta$ ($\mu$rad)}
\end{minipage}
\vskip -2pt
\caption{\label{fig:ideal-plots} Normalized flux for an ideal CCR (WFE = 0 nm,  \(\rho = 1\)). Solid lines correspond to \(\lambda = 532\) nm, while dashed lines correspond to \(\lambda = 1064\) nm. Left is linear scale; right is log scale, focusing on the region with $\alpha\in[3.8,7.5]\,\mu$rad. From top to bottom, the curves represent CCRs with diameters ranging from 110 mm to 80 mm. Compared to Fig.~\ref{fig:Airy-zeros-D}, this figure also shows that the flux scales with aperture size as $\propto D^4$, see (\ref{eq:photon-return-alpha2}). Within the typical velocity-aberration range $\alpha \in [3.8,7.5]\,\mu\mathrm{rad}$, 
smaller apertures outperform larger ones at $532\,\mathrm{nm}$,  whereas at $1064\,\mathrm{nm}$, larger apertures exhibit higher on-axis flux.}
\end{figure}

First, we consider an idealized scenario where the CCRs exhibit no wavefront distortion (\(\mathrm{WFE} = 0\)) and perfect reflectivity (\(\rho = 1\)), isolating the effects of diffraction and velocity aberration. This serves as a fundamental case for assessing performance degradation due to real-world optical imperfections.

Figure~\ref{fig:ideal-plots} illustrates the normalized return flux for ideal CCRs, demonstrating that at 532 nm, single CCRs with diameters of 80–110 mm experience significant flux loss as aberration increases. The flux diminishes rapidly as the aberration approaches the boundary of the main diffraction lobe, nearly vanishing for offsets exceeding \(\sim 5\)–\(7\) \(\mu\)rad. At 1064 nm, the broader diffraction pattern results in a slower flux decline, maintaining detectable photon return even at the largest aberrations, relevant to LLR. 

These results confirm that velocity aberration can significantly degrade photon return. To mitigate this, one can either introduce controlled dihedral angle offsets to modify the far-field diffraction pattern and distribute photons into the second Airy lobe, thereby maintaining LLR operations at reduced flux levels~\cite{Turyshev-etal:2013}, or shift LLR operations to longer wavelengths (e.g., 1064 nm \cite{Turyshev:CW-LLR:2025}), where diffraction effects for majority of LLR stations are less severe.

As shown in Figure~\ref{fig:ideal-plots}, at $532\,\mathrm{nm}$ the comparatively narrow diffraction envelope is highly sensitive to moderate angular velocity aberration offsets in the $\alpha\approx3.9\text{--}7.3\,\mu\mathrm{rad}$ range (see Table~\ref{tab:velocity-aberration-realistic}), which disproportionately penalizes larger apertures. Consequently, smaller apertures at $532\,\mathrm{nm}$ deliver higher on-axis flux under these typical velocity-aberration conditions, favoring a CCR design with reduced diameter to maintain robust returns.

Conversely, at $1064\,\mathrm{nm}$ the broader main lobe tolerates the same angular offsets more effectively, allowing larger apertures to retain a greater fraction of their central intensity. A CCR optimized for $1064\,\mathrm{nm}$ might therefore employ larger apertures to maximize photon return over that same velocity-aberration range.

\subsection{Performance of Realistic CCR Designs}

\begin{figure}[t!]
\begin{minipage}[b]{.48\linewidth}
\rotatebox{90}{\hskip 30pt  Normalized Intensity}
\includegraphics[width=0.95\linewidth]{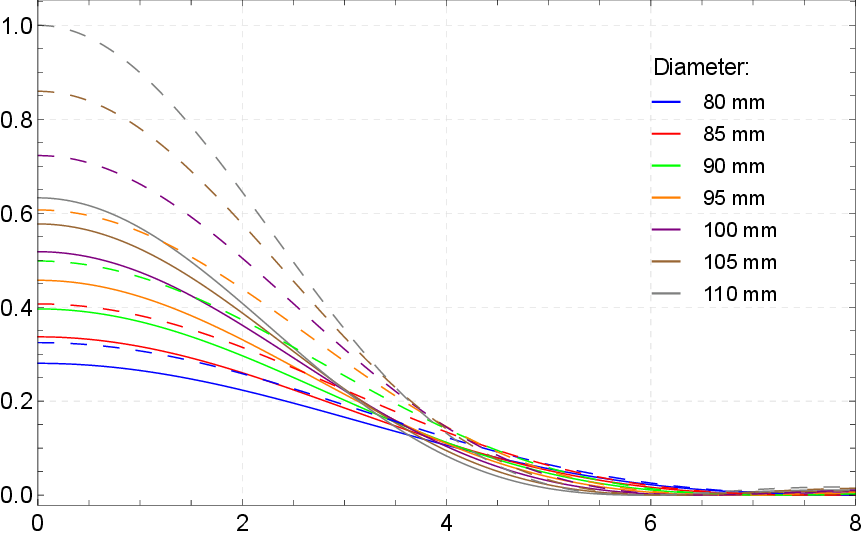}
\rotatebox{0}{\hskip 36pt  Angular displacement $\theta$ ($\mu$rad)}
\end{minipage}
~\,
\begin{minipage}[b]{.50\linewidth}
\rotatebox{90}{\hskip 30pt  Normalized Intensity}
\includegraphics[width=0.95\linewidth]{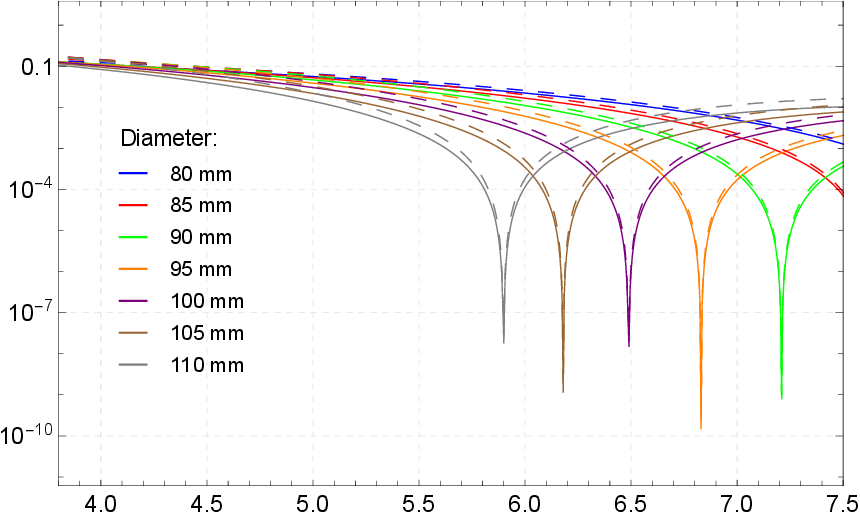}
\rotatebox{0}{\hskip 36pt  Angular displacement $\theta$ ($\mu$rad)}
\end{minipage}
\vskip -2pt
\caption{\label{fig:real-CCR-532nm} Left: Normalized flux anticipated from CCRs at $\lambda=532\,{\rm nm}$  with WFEs from Table~\ref{tab:wfe-strehl-realistic} and reflectivities, \(\rho = 0.92\) for solid and \(\rho = 0.95\) for hollow designs. Solid lines correspond to solid CCRs, while dashed lines are for hollow ones. Right: the same plot, but in logarithmic scale, while zooming in the range with typical values for LLR velocity aberrations at 532 nm. 
}
\begin{minipage}[b]{.485\linewidth}
\rotatebox{90}{\hskip 30pt  Normalized Intensity}
\includegraphics[width=0.95\linewidth]{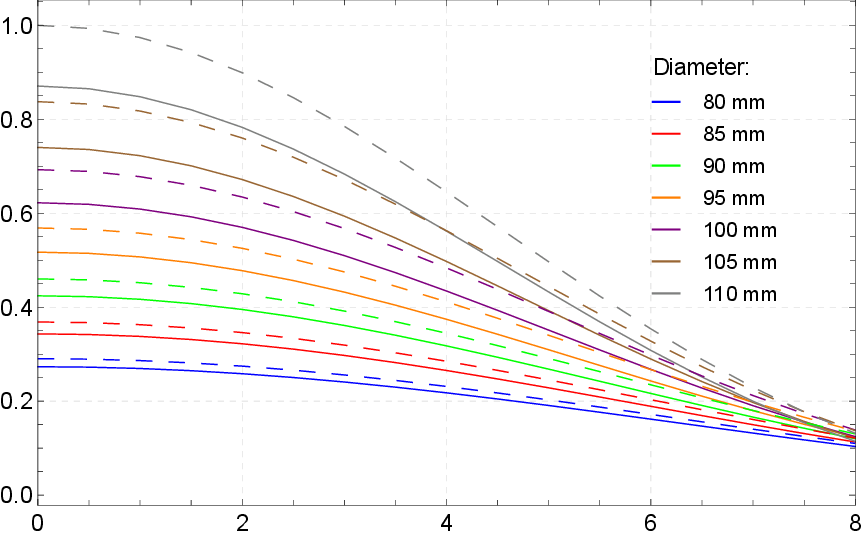}
\rotatebox{0}{\hskip 36pt  Angular displacement $\theta$ ($\mu$rad)}
\end{minipage}
~\,
\begin{minipage}[b]{.49\linewidth}
\rotatebox{90}{\hskip 30pt  Normalized Intensity}
\includegraphics[width=0.95\linewidth]{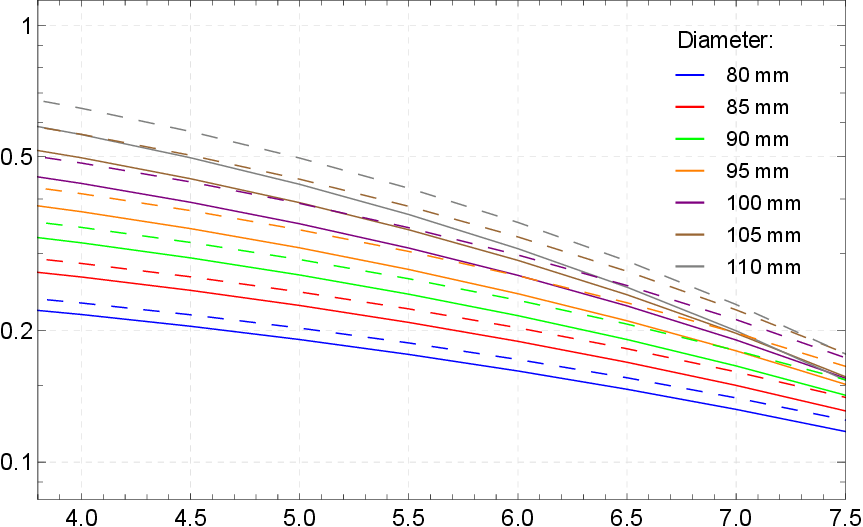}
\rotatebox{0}{\hskip 36pt  Angular displacement $\theta$ ($\mu$rad)}
\end{minipage}
\vskip -2pt
\caption{\label{fig:real-CCR-1064nm} Left: Normalized flux anticipated from CCRs at $\lambda=1064\,{\rm nm}$  with WFEs from Table~\ref{tab:wfe-strehl-realistic} and reflectivities, \(\rho = 0.92\) for solid and \(\rho = 0.95\) for hollow designs. Solid lines correspond to solid CCRs, while dashed lines are for hollow ones. Right: the same plot, while zooming in the range with typical values for LLR velocity aberrations at 1064 nm. 
}
\end{figure}

We explore the performance of realistic CCRs by considering return flux of various designs as a function of  aberration offsets, as given by (\ref{eq:photon-return-alpha2}). Fig.~\ref{fig:real-CCR-1064nm} shows results demonstrating that hollow designs consistently outperform their solid counterparts.  Key performance insights include:
{}
\begin{itemize}
    \item \textit{Aperture Sensitivity to Aberration:} Larger CCRs ($D \geq 100$ mm) deliver maximal flux at small velocity aberrations ($\alpha \leq 2\,\mu\mathrm{rad}$) due to narrower diffraction cones but exhibit sharper flux reductions at larger offsets, favoring mid-sized apertures ($85$–$95\,\mathrm{mm}$) in scenarios with greater velocity aberration.
    
    \item \textit{Impact of Wavefront Errors on Hollow CCR Performance:} Hollow CCRs, with their smaller WFEs ($20$–$48\,\mathrm{nm}$ RMS at $532\,\mathrm{nm}$), consistently achieve better photon return efficiencies relative to their solid counterparts. Crucially, hollow CCRs achieve this performance in a wide range of lunar temperatures \(\Delta T \approx 290^\circ\,{\rm C}\) while reducing total mass by approximately an order of magnitude, providing significant mission-design flexibility. 
    
    \item \textit{Wavelength-Dependent Performance Differences:} The optical performance gap between solid and hollow CCRs diminishes significantly at $1064\,\mathrm{nm}$ due to reduced wavefront sensitivity at longer wavelengths, enhancing the competitiveness of hollow designs at near-infrared wavelengths for new-generation LLR.
\end{itemize}

These results quantitatively confirm the feasibility and practical advantages of hollow CCR designs, particularly in mission architectures where payload mass constraints and thermal stability considerations drive the selection of robust, thermally resilient, and mass-optimized optical retroreflectors. 

Modern bonding methods (e.g.\ brazing of SiC mirrors), advanced mirror polishing, and precision alignment
can mitigate higher-order aberrations such as astigmatism and trefoil. Although hollow CCRs exhibit a slightly smaller acceptance angle \cite{Degnan:2023}, the overall optical performance—whether for solid or hollow—is  determined by the type of a cube, materials and bonding techniques used as well as  the precision of the overall manufacturing process. 

In summary, these tables affirm that hollow SiC CCRs present a robust, lightweight option for high-precision LLR, retaining most of the on-axis flux while drastically reducing payload mass. An informed selection of diameter and wavefront quality can optimize photon return under realistic velocity aberration offsets and harsh lunar thermal cycles.

\subsection{Technical Discussion and Conclusions}

Our simulations reveal several critical trends that inform the design trade-offs for lunar CCRs. In particular, the interplay among velocity aberration, wavelength-dependent WFEs, and the mass and thermal performance of the reflector designs plays a critical role in determining the overall photon return. Key findings include:

\begin{itemize}
  \item \emph{Velocity aberration vs aperture diameter:} For 532 nm, at small angular offsets (\(\alpha \approx 0\)–\(2\,\mu\mathrm{rad}\)), larger CCR apertures (e.g., 100–105\,mm) yield high on-axis flux due to their narrow diffraction patterns. However, as the angular offset increases (to 4, 6, or 8\,\(\mu\mathrm{rad}\)), even moderate velocity-induced misalignments cause a sharp reduction in the main-lobe intensity of large apertures. In these conditions, smaller apertures (80–90\,mm) provide a broader main lobe that maintains a higher net photon return under misalignment. At 1064 nm, impact of aberration on LLR operations is also present, but not as significant as for 532 nm (Figures~\ref{fig:real-CCR-532nm} and \ref{fig:real-CCR-1064nm}). 
  
  \item \emph{Wavefront errors at two wavelengths:}  The same absolute WFE (measured in nm) has a more severe impact at \(\lambda=532\,\mathrm{nm}\) than at \(\lambda=1064\,\mathrm{nm}\), as the error constitutes a larger fraction of the wavelength at 532\,nm. Because a fixed WFEs error corresponds to a smaller phase shift at longer wavelengths, hollow CCRs—with their higher absolute errors—perform more favorably in the near-IR, where the relative phase disturbance is diminished.
  
  \item \emph{Mass advantages of hollow SiC CCRs:}  Hollow CCRs made from SiC offer a substantial reduction in mass---typically an order of magnitude lower mass than solid fused-silica prisms ($\approx$ \(0.18\,\mathrm{kg}\) versus \(2.00\,\mathrm{kg}\) optics-only at $D = 100$ mm; assembled $\sim0.4-0.5$ kg vs. $2.0-2.5$ kg), see Table~\ref{tab:ccr_masses}. Despite exhibiting RMS wavefront errors up to 80\,nm, hollow CCRs retain roughly 80–95\% of the on-axis photon return, particularly at 1064\,nm. This mass efficiency is a decisive advantage for missions requiring multiple retroreflectors or operating under stringent payload constraints.
    
 \item \emph{Thermal-lensing \& mechanical considerations:}
  High thermal conductivity ($120$--$270\,\mathrm{W\,m^{-1}K^{-1}}$) and moderate CTE ($\sim2\times10^{-6}\,\mathrm{K}^{-1}$) in SiC substantially limit thermal gradients, mitigating thermally induced wavefront distortions. In addition, the open geometry of a hollow CCR obviates bulk refractive lensing, thus preserving near-diffraction-limited performance even through lunar day--night temperature swings of $\Delta T\sim290^\circ\mathrm{C}$.
\end{itemize}

In summary, hollow SiC CCRs constitute a robust alternative to conventional solid fused-silica prisms for LLR, offering comparable or superior photon return—particularly at $1064\,\mathrm{nm}$—while significantly reducing payload mass and mitigating thermal-lensing concerns. Recent progress in SiC mirror fabrication, sub-arcsecond interferometric alignment, and precision bonding has enabled these hollow designs to achieve the  WFEs needed for sub-mm-scale LLR. Moreover, the choice of operating wavelength fundamentally shapes the trade‐off among aperture diameter, wavefront stability, and mission‐specific payload constraints. At $532\,\mathrm{nm}$, smaller diameters typically outperform larger ones under typical velocity aberrations (see Table~\ref{tab:velocity-aberration-realistic}), whereas at $1064\,\mathrm{nm}$ larger apertures can fully exploit the broader diffraction lobe to sustain higher flux. Collectively, these results provide a framework for selecting CCR designs that balance optical throughput, thermal-mechanical resilience, and mass limits under realistic lunar conditions.

\section{Hollow CCR Assembly  and Qualification Process}
\label{sec:hollowCCR_Assembly}

The stringent optical performance, mechanical robustness, and long-term environmental stability required for hollow CCRs deployed in LLR demand rigorous, precisely controlled assembly and comprehensive qualification processes. Achieving wavefront errors at the level of $\lambda/15$ ($\sim35\,\text{nm}$ RMS at $532\,\text{nm}$), stable dihedral angles ($90^\circ \pm 0.2''$), and resilience to severe lunar thermal cycling ($-170^\circ\text{C}$ to $+120^\circ\text{C}$), ultrahigh vacuum exposure ($<10^{-10}\,\text{Torr}$), micrometeoroid impacts, intense solar radiation environments, and launch-induced vibrational stresses (up to $20\,g_{\text{RMS}}$) necessitates advanced substrate-specific bonding techniques, ultra-precision alignment methods, and thorough qualification protocols. In this section, we describe these critical processes, quantitatively compare optimized bonding approaches for Zerodur, fused silica, and SiC substrates, and summarize modern precision-assembly methodologies employed to ensure reliable, sustained CCR operation in the challenging lunar environment.

\subsection{Mirror Bonding Techniques for Hollow CCR Assemblies}

Achieving and maintaining the optical precision and long-term stability of hollow CCRs for LLR critically depends on substrate-specific mirror bonding techniques \cite{Preston2013}. These demanding conditions of lunar environment necessitate careful quantitative optimization of bonding methods tailored specifically for Zerodur, fused silica, and SiC substrates.

\subsubsection{Zerodur Mirror Bonding}

Zerodur (lithium aluminosilicate glass-ceramic), with ultra-low thermal expansion ($\alpha_{\text{CTE}}=0.05\times10^{-6}\,\text{K}^{-1}$) and moderate stiffness ($E\approx90\,\text{GPa}$), demands bonding methods that minimally influence dimensional stability and wavefront accuracy:

\begin{itemize}
\item \textit{Optical Contacting:} Surfaces polished to $<1\,\text{nm}$ RMS roughness achieve molecular-level bonding via intermolecular and hydrogen forces. This method introduces negligible stress-induced wavefront distortion ($<0.5\,\text{nm}$ RMS), offers bond strengths of $0.2$--$0.4\,\text{MPa}$, and ensures dimensional stability from $-150^\circ\text{C}$ to $+120^\circ\text{C}$.

\item \textit{Low-Outgassing Epoxy Bonding:} Aerospace-qualified epoxies (e.g., Hysol EA 9313) provide robust mechanical joints with bond strengths of $10$--$15\,\text{MPa}$, but introduce modest wavefront distortion ($5$--$10\,\text{nm}$ RMS) due to thicker bond layers ($50$--$150\,\mu\text{m}$). Rigorous vacuum bake-out ensures outgassing remains below $10^{-6}\,\text{g\,cm}^{-2}\text{s}^{-1}$.
\end{itemize}

\subsubsection{Fused Silica Mirror Bonding}

Fused silica, exhibiting moderate stiffness ($E=72\,\text{GPa}$) and low thermal expansion ($\alpha_{\text{CTE}}=0.55\times10^{-6}\,\text{K}^{-1}$), benefits from chemically stable, minimally invasive bonding techniques:

\begin{itemize}
\item \textit{Hydroxide-Catalysis (Silicate) Bonding:} Employing sodium silicate solutions, this method achieves chemically stable bond thicknesses around $100\,\text{nm}$, with bond strengths of $5$--$15\,\text{MPa}$. Resultant wavefront distortion is typically below $1\,\text{nm}$ RMS, maintaining stability from $-180^\circ\text{C}$ to $+200^\circ\text{C}$ with ultralow outgassing.

\item \textit{Optical Contacting:} Similarly effective, this method provides bond strengths of $0.1$--$0.3\,\text{MPa}$ with minimal wavefront distortion ($<1\,\text{nm}$ RMS). Operational temperature stability spans from $-150^\circ\text{C}$ to $+120^\circ\text{C}$.
\end{itemize}

\subsubsection{Silicon Carbide (SiC) Mirror Bonding}

SiC, characterized by high stiffness ($E\approx410\,\text{GPa}$), superior thermal conductivity ($120$--$270\,\text{W\,m}^{-1}\text{K}^{-1}$), and moderate thermal expansion ($\alpha_{\text{CTE}}=2.2\times10^{-6}\,\text{K}^{-1}$), requires bonding methods capable of withstanding thermal and mechanical stresses:
{}
\begin{itemize}
\item \textit{Active-Metal Brazing (Ti-Ag-Cu alloys):} This robust metallurgical bonding approach yields shear strengths exceeding $50\,\text{MPa}$ with minimal wavefront distortion (typically $2$--$5\,\text{nm}$ RMS). Brazing layers of $50$--$100\,\mu\text{m}$ thickness closely match SiC’s thermal expansion, preserving operational stability from $-180^\circ\text{C}$ to $+250^\circ\text{C}$.

\item \textit{Epoxy Adhesive Bonding:} Aerospace-qualified epoxies (e.g., EPO-TEK 353ND) provide simpler manufacturing processes and achieve bond strengths of $15$--$25\,\text{MPa}$, but introduce moderate wavefront distortion ($5$--$10\,\text{nm}$ RMS). Vacuum bake-out procedures are strictly mandated, achieving outgassing below $10^{-6}\,\text{g\,cm}^{-2}\text{s}^{-1}$.
\end{itemize}

\begin{table}[ht!]
\centering
\caption{Quantitative Comparison of Bonding Techniques for Hollow CCR Mirrors}
\label{tab:bonding-summary}
\renewcommand{\arraystretch}{1.0}
\begin{tabular}{|l|c|c|c|}
\hline
Parameter & Zerodur & Fused Silica & Silicon Carbide (SiC)\\
\hline\hline
Bonding Techniques & Optical/Epoxy & Optical/Hydroxide & Brazing/Epoxy\\
Mirror Thickness & $5$--$8\,\text{mm}$ & $5$--$8\,\text{mm}$ & $3$--$5\,\text{mm}$\\
Bond Strength & $0.2$--$15\,\text{MPa}$ & $0.1$--$15\,\text{MPa}$ & $15$--$50\,\text{MPa}$\\
Thermal Range & $-150^\circ\text{C}$ to $+120^\circ\text{C}$ & $-180^\circ\text{C}$ to $+200^\circ\text{C}$ & $-180^\circ\text{C}$ to $+250^\circ\text{C}$\\
Wavefront RMS & $<0.5$--$10\,\text{nm}$ & $<1\,\text{nm}$ & $2$--$10\,\text{nm}$\\
Bond Thickness & $<10\,\text{nm}$ (optical) & $\sim100\,\text{nm}$ (hydroxide) & $50$--$100\,\mu\text{m}$ (brazing)\\
CTE Mismatch & Minimal & Minimal & Moderate\\
Outgassing & Ultra-low & Ultra-low & Very low (post-bake)\\
Dihedral Stability & $\leq0.2''$ & $\leq0.2''$ & $\leq0.2''$\\
\hline
\end{tabular}
\end{table}

\subsection{Precision Assembly Techniques and Quality Assurance}

Achieving the optical precision and structural stability required by hollow CCR assemblies for LLR necessitates rigorous adherence to advanced precision assembly methods, optimized bonding techniques, and comprehensive qualification procedures. Due to stringent optical requirements—typically WFEs $\leq\lambda/15$ at $532\,\text{nm}$ ($\sim35\,\text{nm}$ RMS)—these assemblies demand exceptional dimensional precision, long-term dihedral angle stability ($90^\circ \pm 0.2''$), and resilience to severe lunar environmental conditions including thermal cycles ($-170^\circ\text{C}$ to $+120^\circ\text{C}$), vacuum exposure ($<10^{-9}\,\text{Torr}$), radiation, micrometeoroid impacts, and launch-induced mechanical stresses (vibration levels up to $20\,g_{\text{RMS}}$).

\subsubsection{Ultra-Precision Surface Polishing and Metrology}

Mirror substrates (SiC, fused silica, Zerodur) undergo deterministic polishing processes utilizing magnetorheological finishing (MRF) or ion-beam figuring (IBF), achieving surface figure accuracy $<1\,\text{nm}$ RMS and micro-roughness $<0.5\,\text{nm}$ RMS. Surface quality and wavefront accuracy are verified interferometrically using high-resolution phase-shifting interferometry at  $\lambda=532\,\text{nm}$ and $\lambda=632.8\,\text{nm}$, ensuring minimal post-assembly optical distortion.

\subsubsection{Precision Dihedral Angle Alignment and Verification}

CCR optical performance critically depends on maintaining precise dihedral angles ($90^\circ\pm0.2''$ or $\pm1\,\mu\text{rad}$). Alignment is executed via robotic hexapod positioning stages integrated with real-time interferometric feedback. Achieved positional accuracies routinely reach $<0.5\,\mu\text{m}$ translationally and $<0.05''$ ($0.25\,\mu\text{rad}$) rotationally. Post-bonding interferometric verification ensures consistent dihedral angle precision is maintained after exposure to thermal cycling and simulated launch vibrations.

\subsubsection{Optimized Bonding Techniques and Thermal Management}

Substrate-appropriate bonding methods are meticulously selected to minimize induced wavefront distortion and thermal mismatch stresses:
\begin{itemize}
    \item \textit{Optical contacting} (Zerodur, fused silica): Ultra-thin, adhesive-free bonding ($<10\,\text{nm}$), resulting in negligible wavefront distortion ($<1\,\text{nm}$ RMS), bond strengths $0.2$--$0.4\,\text{MPa}$, and exceptional thermal stability.
    \item \textit{Hydroxide-catalysis bonding} (fused silica): Chemically bonded joints $\sim100\,\text{nm}$ thick, achieving bond strengths $5$--$15\,\text{MPa}$, wavefront distortions $<1\,\text{nm}$ RMS, and operational stability over extensive lunar thermal cycles ($-170^\circ\text{C}$ to $+120^\circ\text{C}$).
    \item \textit{Active-metal brazing} (SiC substrates): Employing AgCuTi alloys processed at $800$--$950^\circ\text{C}$, producing bond strengths exceeding $20\,\text{MPa}$, bond layers $50$--$100\,\mu\text{m}$, and minimal thermal mismatch (CTE difference $<0.5\times10^{-6}\,\text{K}^{-1}$), yielding wavefront distortions of $2$--$5\,\text{nm}$ RMS.
\end{itemize}

Advanced thermal management techniques, such as high-emissivity coatings, tailored thermal interface materials, and conductive braze layers, limit mirror thermal gradients to $<5^\circ\text{C}$, further ensuring minimal optical distortion during lunar operation.

\subsubsection{Comprehensive Thermal and Mechanical Qualification}

CCRs must undergo rigorous environmental qualification testing to verify structural and optical robustness under lunar and launch conditions:
\begin{itemize}
    \item \textit{Thermal vacuum cycling}: $>500$ cycles spanning $-170^\circ\text{C}$ to $+120^\circ\text{C}$, demonstrating WFEs within $5\,\text{nm}$ RMS.
    \item \textit{Vibration and shock testing}: Random vibration tests conforming to NASA-GEVS specifications (up to $20\,g_{\text{RMS}}$) ensure structural integrity, with dihedral angle deviations maintained below $0.2''$ post-test.
    \item \textit{Finite-element modeling (FEM)}: Comprehensive FEM analysis quantifies mechanical stress distribution, thermally induced deformation, and optical distortions, providing predictive validation of performance under realistic launch and lunar conditions.
\end{itemize}

\subsubsection{Long-term Environmental Stability and Outgassing Control}

Extended vacuum bake-outs ($>$500 hours at pressures $<10^{-8}\,\text{Torr}$) ensure ultra-low outgassing rates ($<10^{-6}\,\text{g\,cm}^{-2}\text{s}^{-1}$). Reflective coatings, comprising Ag or Au layers with dielectric protective multilayers, undergo radiation exposure (proton/electron flux) and accelerated thermal aging tests, verifying sustained reflectivity ($\rho>0.98$) over operational lifetimes exceeding 30 years. Dust contamination mitigation employs electrostatic dissipative coatings, reducing lunar dust particle adhesion forces below $1\,\mu\text{N}$ per particle.

\subsubsection{Quantitative Technical Summary}

Technical capabilities achieved using current precision assembly and qualification techniques for hollow CCRs are quantitatively summarized in Table~\ref{tab:precision-assembly-summary}. Collectively, these state-of-the-art assembly practices, material-specific bonding techniques, and comprehensive qualification strategies ensure the CCRs achieve the stringent optical, mechanical, and environmental performance metrics essential for sustained, high-accuracy LLR.

\begin{table}[ht!]
\centering
\caption{Quantitative Summary of Precision Assembly and Qualification for Hollow CCRs}
\label{tab:precision-assembly-summary}
\renewcommand{\arraystretch}{1.0}
\begin{tabular}{|l| c|}
\hline
Parameter & Achieved Specification \\
\hline\hline
Mirror surface figure (post-polishing) & $<1\,\text{nm}$ RMS \\
Surface micro-roughness & $<0.5\,\text{nm}$ RMS \\
Translational alignment accuracy & $<0.5\,\mu\text{m}$ \\
Angular alignment accuracy & $<0.05''$ ($<0.25\,\mu\text{rad}$) \\
Dihedral angle precision & $90^\circ\pm0.2''$ ($\sim\pm1\,\mu\text{rad}$) \\
Wavefront error (assembled, $D=100\,\text{mm}$) & $25$--$35\,\text{nm}$ RMS ($\lambda/15$ at $532\,\text{nm}$) \\
Wavefront drift after thermal cycling & $<25\,\text{nm}$ RMS \\
Wavefront drift after vibration (20\,$g_{\text{RMS}}$) & $<25\,\text{nm}$ RMS \\
Bond thickness & Optical/Hydroxide ($<100\,\text{nm}$), Brazing ($50$--$100\,\mu\text{m}$) \\
Bond strength & Optical ($0.2$--$0.4\,\text{MPa}$), Brazing ($>20\,\text{MPa}$) \\
Operational thermal range & $-170^\circ\text{C}$ to $+120^\circ\text{C}$ \\
Outgassing rate & $<10^{-6}\,\text{g\,cm}^{-2}\text{s}^{-1}$ \\
Reflectivity stability (mission life $>$30 years) & $>0.98$ (Ag/Au + dielectric overcoat) \\
Dust adhesion mitigation & Electrostatic dissipative coatings ($<1\,\mu\text{N}$ per particle) \\
\hline
\end{tabular}
\end{table}

\subsection{Flight heritage and use of hollow CCRs}
\label{sec:heritage}
To date, lunar LLR retroreflectors have used \emph{solid} fused-silica prisms—the Apollo and Lunokhod arrays and, more recently, the 100\,mm Next-Generation Lunar Retroreflector (NGLR)~\cite{Currie_etal_2011,Turyshev-etal:2013, Williams-etal:2023}. Mirror-based (hollow) trihedrals are mature for terrestrial metrology and astronomy, but their flight heritage for LLR is limited. This paper provides a wave-optics basis and design guidance quantifying the mass/thermal advantages and performance margins of hollow SiC CCRs for future lunar deployments.

\subsection{Mechanical Stability and Thermal Expansion in Hollow CCRs}
\label{subsec:hollow-thermal-flexure}

Unlike solid CCRs, which experience refractive distortions due to temperature-dependent variations in the silica refractive index, hollow CCRs are susceptible to mechanical flexure caused by differential thermal expansion between the mirrors and their supporting structures.

\subsubsection{Thermal Expansion Effects on Optical Stability}

Differential thermal expansion across bonded mirror assemblies can lead to dihedral angle distortions, introducing wavefront errors exceeding 50 nm RMS. Table~\ref{tab:cte-materials} summarizes the CTEs for key materials used in hollow CCRs.

\begin{table}[ht!]
\centering
\caption{Coefficient of thermal expansion (CTE) for materials used in hollow CCR mirrors.}
\label{tab:cte-materials}
\renewcommand{\arraystretch}{1.0}
\begin{tabular}{|c|c|c|}
\hline
Material & CTE (\(\times 10^{-6}\) K\(^{-1}\)) & Young’s Modulus (GPa) \\
\hline\hline
Zerodur & 0.05 & 90 \\
Fused Silica & 0.55 & 72 \\
Silicon Carbide (SiC) & 2.2 & 410 \\
Titanium Bonding Layer & 8.5 & 116 \\
\hline
\end{tabular}
\end{table}

The thermal expansion-induced deformation of a 100-mm SiC mirror under a lunar diurnal cycle (\(\Delta T = 290\) K) is:
$
    \Delta L_{\text{SiC}} = 100\,\text{mm} \times (2.2 \times 10^{-6}\,\text{K}^{-1}) \times 290\,\text{K} = 0.64\,\text{mm}.
$
For Zerodur:
$
    \Delta L_{\text{Zerodur}} = 100\,\text{mm} \times (0.05 \times 10^{-6}\,\text{K}^{-1}) \times 290\,\text{K} = 0.0145\,\text{mm}.
$
These deformations can disrupt dihedral angles, introduce WFEs $\sim$ 50 nm RMS, and degrade photon return efficiency. This highlights the need for precision bonding techniques that mitigate stress-induced flexure.

\subsubsection{Finite Element Analysis (FEA) of Hollow CCR Stability}

An FEA study was conducted for a 100-mm hollow CCR with SiC mirrors bonded to a titanium support structure to analyze thermal stress-induced flexure. Results are summarized in Table~\ref{tab:hollow-ccr-flexure}.

\begin{table}[ht!]
\centering
\caption{Thermally induced flexure in a 100-mm hollow CCR under lunar diurnal cycling.}
\label{tab:hollow-ccr-flexure}
\renewcommand{\arraystretch}{1.0}
\begin{tabular}{|c|c|c|c|}
\hline
Material Combination & Max Flexure (\(\mu\)rad) & Wavefront Error (nm RMS) & Impact on Strehl Ratio \\
\hline\hline
SiC-Titanium Bonding  & 2.5  & 45  & 0.85 \\
SiC-Zerodur Bonding  & 1.1  & 25  & 0.92 \\
SiC-Fused Silica Bonding  & 0.9  & 18  & 0.96 \\
\hline
\end{tabular}
\end{table}

SiC-Ti assemblies experience the largest flexure, leading to a wavefront degradation of 45 nm RMS and reducing the Strehl ratio to 0.85. In contrast, SiC-Zerodur and SiC-Fused Silica CCRs exhibit better thermal stability, with WFEs below 25 nm, ensuring higher photon return efficiency.

\subsubsection{Mitigation Strategies}

To reduce the thermal distortion effects described above, several approaches
can be employed:

\begin{itemize}
    \item \textit{Use Zerodur or fused-silica mounts rather than titanium.}
    These ultra-low expansion ceramics and glasses exhibit minimal  CTE mismatch when bonded to SiC  or other low-CTE mirror substrates. Consequently, the overall mount–mirror  assembly is less susceptible to thermally induced stress and dihedral-angle  drifts. Titanium, by contrast, has a significantly higher CTE and can impose larger differential strains during lunar day–night temperature swings (up to 290\,$^\circ$C range), leading to wavefront degradation.

    \item \textit{Use low-expansion adhesives, e.g., hydroxide-catalysis bonding.}
 Bonding methods that produce very thin, chemically stable joints—on the  order of 10s to 100s of nanometers—greatly reduce the risk of bulk adhesive expansion or shrinkage. Hydroxide-catalysis (or ``silicate'') bonding, for instance, forms near-monolithic bonds between polished surfaces and can achieve bond strengths of several MPa while introducing  only negligible ($\sim$1\,nm) wavefront distortion. This helps keep the mirror edges accurately aligned, even under wide thermal swings.

    \item \textit{Apply high-conductivity coatings (e.g.\ gold or silver) to homogenize temperature distributions.} By increasing the reflectivity and thermal conductivity of the mirror surfaces, these metallic overcoats (often combined with protective dielectric layers) can reduce local hot spots or thermal gradients arising from direct solar illumination. A more uniform mirror temperature profile translates into less mechanical flexure, thus preserving wavefront quality and dihedral-angle stability.

\end{itemize}

Collectively, these measures mitigate thermally induced strain and reduce mechanical deformations that would otherwise shift the far-field diffraction pattern or degrade wavefront fidelity. As a result, hollow CCRs assembled
with careful material selection and bonding practices are better able to maintain long-term optical stability in the extreme lunar environment, where day–night cycles can exceed a 290\,$^\circ$C temperature range.

\section{Optimal CCR Design for LLR}
\label{sec:optimal_ccr_design}

\subsection{Materials and Geometry}

This section provides an examination of the design trade-offs involved in selecting a CCR capable of achieving a sub-mm LLR performance. The wave-optics simulations and thermal-mechanical studies in Secs.~\ref{sec:thermalWFE}, \ref{sec:simulation} and \ref{sec:simulation-results} investigate a range of \(\mathrm{CCR}\) apertures 80--110\,mm  under two principal constraints: a \(\,\sim290^\circ\mathrm{C}\) lunar day--night thermal cycle and velocity aberrations on the order of \(4\text{--}7\,\mu\mathrm{rad}\). The subsequent subsections discuss how these thermal, optical, and mechanical considerations jointly inform the choice of aperture diameter and retroreflector geometry.  

\subsubsection{Aperture Diameter and Velocity Aberration Tolerance}

Numerical simulations (Secs.~\ref{sec:simulation}--\ref{sec:simulation-results}) indicate that \(\mathrm{CCR}\) apertures below \(\sim80\,\mathrm{mm}\) generally fail to gather enough photons for sub-millimeter \(\mathrm{LLR}\) precision. On the other hand, larger diameters (exceeding \(\sim110\,\mathrm{mm}\)) produce relatively narrow main diffraction lobes, making them highly susceptible to modest velocity offsets (\(\alpha\approx4\text{--}7\,\mu\mathrm{rad}\)). Depending on the wavelength and specific alignment, these offsets can lead to on-axis flux drops  \(\sim40\%\text{--}50\%\) or higher when a significant portion of the main lobe is displaced outside the receive aperture. The severity of such losses is often worse at \(\lambda=532\,\mathrm{nm}\), since its diffraction disk is inherently smaller than at \(\lambda=1064\,\mathrm{nm}\).

An intermediate diameter near \(100\,\mathrm{mm}\) balances these trade-offs across both wavelengths. It remains large enough to provide robust collection efficiency for sub-mm ranging, while not so large that moderate velocity aberrations severely reduce the central lobe intensity. Thus, a \(\,100\,\mathrm{mm}\)-aperture \(\mathrm{CCR}\) tends to preserve adequate on-axis flux at typical lunar velocity offsets, improving the likelihood of sustained high-precision LLR over the full mission timeline.

\subsubsection{Lunar Thermal Cycle  and Wavefront Deformations}

Fused silica possesses a modest CTE and a relatively low thermal conductivity, see Table~\ref{tab:ccr-parameters-multi}. During a diurnal lunar temperature swing of  
\(\sim 290^\circ\mathrm{C}\), these properties cause significant gradients within a solid fused-silica prism, giving rise to refractive-index variations that elevate WFEs to about \(50\text{--}70\,\mathrm{nm\,RMS}\) for a \(\sim 100\,\mathrm{mm}\) aperture, see Table~\ref{tab:wfe_ccrs}.

In contrast, hollow SiC CCRs provide substantially higher thermal conductivity and a minimal bulk volume, confining internal \(\Delta T\) to only a few degrees. This design typically restricts realistic WFEs to below \(25\,\mathrm{nm\,RMS}\)  over the full \(\sim 290^\circ\mathrm{C}\) cycle, thereby 
maintaining near-diffraction-limited performance through both lunar day and night.

\subsubsection{Reflectivity: TIR vs.\ Mirror-Based Reflection}

Although TIR in a fused-silica prism can provide per-bounce reflectivities of \(\gtrsim0.99\), Fresnel losses at the entrance and exit faces typically reduce the overall reflectivity to around \(\sim0.92\text{--}0.93\). In contrast, hollow SiC CCRs rely on three external mirror reflections, each with \(\rho \approx 0.98\) when using modern protected-silver (\(\mathrm{Ag + SiO_2/TiO_2}\)) coatings, yielding a net reflectivity of \((0.98)^3 \approx 0.94\). Consequently, both designs achieve broadly similar performance in the \(\sim0.92\text{--}0.95\) range. Reflectivity alone, therefore, is not the deciding criterion for geometry selection. Instead, factors such as overall mass, thermal and mechanical stability generally govern the optimal choice of CCR architecture.

\subsubsection{Mass Budget and Manufacturing Feasibility}

For a \(100\,\mathrm{mm}\)-aperture CCR, monolithic fused-silica designs typically weigh between \(1.8\) and \(2.4\,\mathrm{kg}\), depending on prism height and mounting details. In contrast, hollow SiC CCRs of comparable aperture weigh only about \(0.4\text{--}0.5\,\mathrm{kg}\), representing an approximate \(80\%\) mass reduction. This lighter mass is especially advantageous in lunar robotic missions, where stringent payload constraints heavily influence both mission design and capabilities.

Fused-silica CCRs can be polished to high optical standards and assembled using hydroxide-catalysis bonding or optical contacting. These processes reliably attain dihedral-angle tolerances of \(\leq 0.2''\) and WFEs under \(25\,\mathrm{nm\,RMS}\). However, working with large fused-silica blanks increases fabrication complexity and cost, and repeated diurnal thermal swings on the Moon may cause dihedral-angle drifts of \(\sim 0.5''\). Over time, such angular shifts degrade photon-return efficiency and compromise long-term LLR precision.

By comparison, hollow SiC CCRs rely on individually polished mirror facets joined via active-metal brazing, diffusion bonding, or similarly specialized ceramic-joining methods—well-established approaches for SiC. These techniques achieve optical precisions on par with fused silica. Moreover, the higher thermal conductivity and reduced bulk of SiC structures mitigate thermo-mechanical stresses, thus delivering superior dimensional stability and enhanced optical performance across the harsh lunar day--night temperature extremes.

\subsubsection{Quantitative Summary}

Table~\ref{tab:ccr_SiC_vs_silica} summarizes key performance metrics for a \(100\,\mathrm{mm}\)-aperture CCR operating under the demanding lunar thermal environment of \(\sim290^\circ\mathrm{C}\). The hollow SiC CCR consistently exhibits superior thermo-mechanical stability, as evidenced by minimal wavefront distortion (\(\leq5\,\mathrm{nm}\,\mathrm{RMS}\)), and significantly reduced mass (\(0.4\text{--}0.5\,\mathrm{kg}\)). Although total reflectivity values are broadly comparable between solid fused-silica and hollow SiC CCRs, mass constraints, thermal properties, and long-term mechanical stability decisively favor the hollow SiC design. Wave-optics and finite-element analyses presented in earlier sections confirm these advantages, establishing the \(100\,\mathrm{mm}\) hollow SiC CCR as the optimal solution for future sub-mm LLR operations.

\begin{table}[ht!]
\centering
\caption{Quantitative comparison of hollow SiC versus solid fused-silica CCR designs at \(100\,\mathrm{mm}\) aperture, subjected to lunar diurnal temperature cycles (\(\sim290^\circ\mathrm{C}\)).}
\label{tab:ccr_SiC_vs_silica}
\vspace{4pt}
\begin{tabular}{|l|cc|}
\hline
Parameter & Solid Fused Silica & Hollow SiC\\
\hline\hline
Wavefront error & \(50\text{--}70\,\mathrm{nm}\,\mathrm{RMS}\) & \(\leq25\,\mathrm{nm}\,\mathrm{RMS}\)\\
Thermal conductivity & \(\sim1.3\,\mathrm{W\,m^{-1}\,K^{-1}}\) & \(\sim120\,\mathrm{W\,m^{-1}\,K^{-1}}\)\\
Mass & \(2.0\text{--}2.5\,\mathrm{kg}\) & \(0.4\text{--}0.5\,\mathrm{kg}\)\\
Total reflectivity & \(\sim0.92\text{--}0.93\) & \(\sim0.94\text{--}0.95\)\\
Dihedral-angle stability & \(\approx0.5''\) & \(\leq0.2''\)\\
\hline
\end{tabular}
\end{table}

\subsection{Deployment and Acceptance Angles}
\label{sec:ccr-placement}

Achieving sub-mm-level LLR measurements requires a CCR system that preserves high photon returns despite large thermal fluctuations and potential mechanical misalignments on the lunar surface. In our design, advanced hollow SiC CCRs are used to exploit their high thermal conductivity and low mass, thus minimizing thermally induced WFEs.

\subsubsection{Dual-Reflector Configuration and Co-Boresighting}
\label{sec:dual-reflector}

The dual-CCR layout serves three goals: (i) improve \emph{azimuthal robustness} to velocity-aberration/libration by ``clocking'' the two far-field patterns so at least one strong lobe remains near boresight; (ii) enable \emph{differential ranging} across a short, thermally monitored baseline to estimate and remove lander-induced path changes; and (iii) provide \emph{operational redundancy} for long-duration operations.

Two hollow CCRs, each measuring \(100\,\mathrm{mm}\) in diameter, are rigidly mounted on a single platform with a precisely controlled \(0.50\,\mathrm{m}\) baseline. This dual-reflector assembly is co-aligned with the lander's high-gain antenna (HGA), ensuring that upon lunar touchdown, both CCR optical axes remain oriented toward Earth. Given that the HGA maintains pointing accuracy to within \(\pm 10^\circ\), the mounting design keeps residual angular misalignments between the CCR axes and the terrestrial laser line-of-sight to only a few degrees.

FEM predicts that stress-induced tilt of the CCRs remains under \(0.2''\) across thermal excursions of up to \(300\,\mathrm{K}\). Such minimal deformations preserve near-diffraction-limited optical performance and ensure the high photon return necessary for sub-mm-scale precision in LLR.

\subsubsection{Diffraction-Limited Performance and Acceptance Angles}
\label{sec:diffraction}

For a CCR aperture of diameter $D = 100\,\mathrm{mm}$ operating at a wavelength $\lambda = 1064\,\mathrm{nm}$, the far-field diffraction pattern follows the Airy function, with a first-null angular radius given by $\theta_{\mathrm{null}} \approx 1.22\lambda/D \approx 12.97\,\mu\mathrm{rad}$. Velocity aberrations induced by lunar orbital motion ($\sim 1\,\mathrm{km/s}$) and Earth's rotation (233--465 m/s) (see Sec.~\ref{sec:vel-aberr} and Table~\ref{tab:velocity-aberration-realistic}) typically lie within the range of $4$--$8\,\mu\mathrm{rad}$ for lunar observations at elevations down to $\sim 10^\circ$.

To ensure robust LLR performance, the CCR acceptance angle—defined as the angular range over which photon returns exceed $50\%$ of the peak on-axis efficiency—must comfortably exceed these velocity-aberration offsets. In practice, the effective peak cross-section is reduced by an efficiency correction factor, $\eta(\theta_{\text{inc}})$, which accounts for the degradation in performance as the beam incidence angle, $\theta_{\text{inc}}$, increases, see \cite{Degnan:2023}. Specifically, the effective cross-section is given by
{}
\begin{equation}
\sigma_{\text{eff}}(\theta_{\text{inc}})=\eta^2(\theta_{\text{inc}})\,\sigma_{\text{cc}},
\end{equation}
where $\sigma_{\text{cc}}$ is the on-axis cross-section. The efficiency correction factor is defined by
{}
\begin{equation}
\eta(\theta_{\text{inc}})=\frac{2}{\pi}\left[\sin^{-1}\left(\sqrt{1-\tan^2\theta_{\text{ref}}}\,\right)-\sqrt{2}\tan\theta_{\text{ref}}\cos\theta_{\text{inc}}\right],
\end{equation}
with the internal refracted angle, $\theta_{\text{ref}}$, determined by Snell’s law as
{}
\begin{equation}
\theta_{\text{ref}}=\sin^{-1}\left(\frac{\sin\theta_{\text{inc}}}{n}\right),
\end{equation}
and $n$ being the index of refraction of the cube material. 

\begin{figure}[h!]
\centering
\rotatebox{90}{\hskip 18pt  Normalized CCR cross-section}\hskip 6pt
\includegraphics[width=0.45\linewidth]{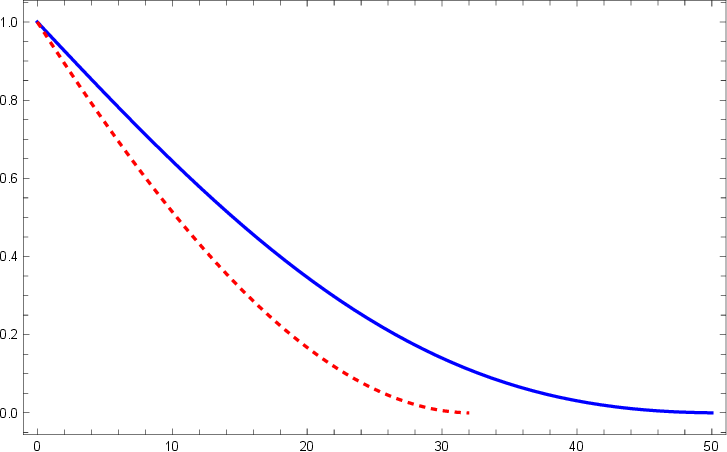}\\
\rotatebox{0}{\hskip 30pt  Incidence angles (degrees)}
\caption{\label{fig:Incidence-angles} The normalized intensity of the returned signal as a function of beam incidence angle differs notably between solid fused-silica ($n=1.455$, solid line) and hollow ($n=1$, dotted line) CCRs. Solid fused-silica CCRs retain 50\% efficiency up to  $\sim13^\circ$ incidence, declining to zero near $45^\circ$. In contrast, hollow CCRs exhibit a narrower acceptance, maintaining 50\% efficiency only up to about $10^\circ$ and reaching zero around $31^\circ$. Consequently, hollow CCRs require more stringent alignment or co-boresighted reflector configurations to sustain robust performance under lunar libration and typical operational misalignments. }
\end{figure}

Figure~\ref{fig:Incidence-angles} plots the correction factor $\eta^2$ for silica ($n=1.455$) and hollow ($n=1$) CCRs. It is evident that hollow CCRs exhibit a narrower angular acceptance compared to solid fused-silica prisms. Specifically, hollow CCR efficiency declines to $50\%$ at an incidence angle of roughly $10^\circ$ and effectively drops to zero at $\sim31^\circ$, while solid fused-silica CCRs retain $50\%$ efficiency out to $\sim13^\circ$ incidence and fully lose efficiency near $45^\circ$. Therefore, hollow SiC CCRs inherently require tighter alignment constraints.  

To mitigate these constraints, we adopt a dual-reflector configuration in which two hollow CCRs are intentionally co-boresighted with small, controlled angular offsets. By carefully selecting these angular offsets, the combined system achieves a broadened effective angular acceptance, thereby maintaining high photon return rates even when lunar librations induce several-degree variations in the lander orientation. Comprehensive wave-optics and finite-element simulations confirm that this co-boresighted arrangement consistently preserves overlap efficiency above the $50\%$ threshold across realistic lunar operational scenarios, ensuring sustained precision for sub-mm LLR \cite{Turyshev:CW-LLR:2025}.

\subsubsection{Differential Ranging and Thermal Expansion Compensation}
\label{sec:differential}

In earlier work~\cite{Turyshev:CW-LLR:2025} km-scale baselines between widely separated CCRs were considered primarily to enable the sensitivity to deep lunar interior with the differential LLR paradigm. Here the fixed baseline is $L=0.50$\,m and targets a different goal: \emph{in-situ compensation} of \emph{local thermo-mechanical expansion} of the lander/fixture at the $10$–$20\,\mu$m level during a session. This shorter baseline keeps both CCRs inside the HGA boresight, simplifies integration, and still provides the needed sensitivity (e.g., for an aluminum structure with $\alpha=2.3\times10^{-5}{\rm K}^{-1}$ over $\Delta T=300\,{\rm K}$, the expected expansion is $\Delta L \simeq 3.45\,{\rm mm}$, which we track and remove well below $0.1\,{\rm mm}$).   

The fixed baseline of approximately $L = 0.50\,\mathrm{m}$ between the two CCRs facilitates differential ranging between the CCRs in the pair, enabling precise in-situ compensation for thermal and mechanical perturbations. Over typical lunar diurnal temperature cycles ($\Delta T \approx 300\,\mathrm{K}$), thermal gradients cause structural deformations that would normally compromise absolute range accuracy. For example, assuming an aluminum support structure with a thermal expansion coefficient $\alpha \approx 2.3\times10^{-5}\,\mathrm{K}^{-1}$, the expected baseline expansion is
\[
   \Delta L \;=\; L\,\alpha\,\Delta T \;\approx\; 0.50\,\mathrm{m}\,\times\,2.3\times10^{-5}\,\mathrm{K}^{-1}\,\times\,300\,\mathrm{K}
   \;\approx\;3.45\,\mathrm{mm}.
\]

Although this magnitude of mechanical expansion significantly exceeds the targeted sub-mm LLR precision, differential measurements of the two nearly-simultaneously returned signals \cite{Turyshev:CW-LLR:2025} enable direct monitoring of baseline changes with resolutions of $\sim 10$--$20\,\mu\mathrm{m}$. This approach effectively reduces thermal baseline drift errors to well below the $0.1\,\mathrm{mm}$ threshold required for sub-mm ranging performance, thus preserving the integrity of high-precision lunar geophysical measurements.

\subsubsection{Aperture Selection, Wavelength Considerations, and Mass Benefits}
\label{sec:aperture}

Wave-optics simulations indicate that CCR apertures larger than approximately $110\,\mathrm{mm}$ become increasingly sensitive to modest velocity aberration offsets ($\sim 4$--$7\,\mu\mathrm{rad}$). These larger apertures produce narrower diffraction lobes, leading to substantial reductions (often exceeding $50\%$) in on-axis photon returns even under small misalignments. Conversely, apertures smaller than about $80\,\mathrm{mm}$ do not provide sufficient photon flux to reliably achieve sub-millimeter ranging precision, especially given realistic observational noise conditions. Hence, a CCR diameter of about $100\,\mathrm{mm}$ emerges as an optimal compromise, maintaining robust photon returns and tolerating typical lunar velocity aberrations, with a diffraction-limited first-null angular radius of approximately $\theta_{\mathrm{null}} \approx 12.97\,\mu\mathrm{rad}$ at $\lambda = 1064\,\mathrm{nm}$.

Employing hollow SiC structures significantly reduces reflector mass to $\sim 0.4$--$0.5\,\mathrm{kg}$ per $100\,\mathrm{mm}$ CCR, compared to solid fused-silica counterparts weighing roughly $2.0$--$2.5\,\mathrm{kg}$. Such mass savings are particularly critical for lunar payloads, where every kilogram is highly constrained and directly affects mission scope and capability. Additionally, operation at a near-infrared wavelength of $1064\,\mathrm{nm}$ improves photon-return performance. The larger diffraction spot at this wavelength provides increased tolerance to velocity aberration, resulting in higher overlap efficiency with ground-based receiving telescopes and mitigating Fresnel and TIR reflection losses typical at shorter wavelengths.

Combining these local measurements with global Earth--Moon data yields a multi-scale determination of lunar librations, interior structure, and fundamental tests of gravitational physics.

\subsection{Summary and Technical Conclusions}
\label{sec:summary}

Under realistic lunar velocity aberrations (approximately $3$--$8\,\mu\mathrm{rad}$), extreme thermal cycles (up to $300\,\mathrm{K}$), and structural distortions, our dual-reflector configuration of two hollow SiC CCRs (each $100\,\mathrm{mm}$ in diameter) reliably achieves sub-millimeter lunar laser ranging precision. The principal technical advantages of this configuration include:

\begin{itemize}
\item \textit{Stable Dual-CCR Baseline:} The $0.50\,\mathrm{m}$ baseline, integrally co-boresighted with the lander’s HGA, maintains robust Earth-pointing accuracy despite typical lander attitude variations.

\item \textit{Optimized Acceptance Angles:} With a diffraction-limited angular half-width of $\sim12.97\,\mu\mathrm{rad}$ at $\lambda = 1064\,\mathrm{nm}$, the CCRs comfortably accommodate typical velocity aberrations and moderate alignment errors encountered on the lunar surface.

\item \textit{High-Resolution Differential Ranging:} Precise differential measurements across the $0.50\,\mathrm{m}$ baseline resolve thermal-induced expansions at the $10$--$20\,\mu\mathrm{m}$ level \cite{Turyshev:CW-LLR:2025}, effectively mitigating positional uncertainties arising from local structural deformation.

\item \textit{Superior Mass Efficiency:} The hollow SiC CCR units (each $0.4$--$0.5\,\mathrm{kg}$) offer substantial mass savings relative to solid fused-silica prisms (each $2.0$--$2.5\,\mathrm{kg}$), a critical advantage for payload-constrained lunar missions.

\item \textit{CCR Network Scalability \& Integration:} Deployment of multiple landers, each equipped with a dual-CCR system, provides a distributed network capable of high-fidelity measurements. This multi-scale architecture significantly improves geophysical investigations, lunar libration measurements, and fundamental gravitational tests over extended mission durations.
\end{itemize}

Collectively, these technical enhancements—large-aperture hollow CCR designs, precise co-boresighting, and differential baseline metrology—address the stringent angular acceptance and thermal stability requirements intrinsic to lunar operations. This optimized system thus provides a robust, high-precision LLR capability suitable for advanced geophysical research and fundamental physics experiments.

\section{Conclusion}
\label{sec:conclusion}

We have developed a comprehensive wave‐optics model for lunar corner‐cube retroreflectors (CCRs) with apertures spanning 80–110\,mm. Our simulation framework integrates rigorous Fraunhofer diffraction theory with spatially varying thermal–mechanical wavefront error (WFE) maps—expressed via Zernike polynomials—and realistic reflectivity values ($\approx\,$0.92 for solid fused‐silica CCRs and 0.95 for hollow silicon carbide (SiC) CCRs). This multi-parameter model quantitatively captures the interplay among aperture size, WFE magnitude, and partial velocity aberration (up to 8\,$\mu$rad), which together determine the photon-return flux—a critical metric for lunar laser ranging (LLR).

Our analysis reveals a pronounced diameter–offset coupling. Under near-ideal wavefront conditions, larger CCRs (e.g., 100–110\,mm) exhibit high on-axis intensity due to their narrow diffraction cones; however, their performance degrades sharply with even moderate angular offsets (4–6\,$\mu$rad), leading to substantial losses in main-lobe overlap. In contrast, mid-sized CCRs (80–90\,mm) possess broader main lobes that are more tolerant of misalignment, thereby maintaining a higher net photon return under realistic operational conditions.

We note that many modern LLR stations, such as APOLLO \cite{Murphy_etal_2008, Battat_etal_2009}, currently operate at 532\,nm. However, our analysis demonstrates a compelling need to transition to 1064\,nm. For the new generation of small, single CCRs \cite{Currie_etal_2011}, operating at 532\,nm pushes the photon return toward the edge of the Airy disk due to velocity aberration, often resulting in many near-zero returns. In contrast, at 1064\,nm the returned signal remains well within the main lobe of the diffraction pattern. Since OCA \cite{Samain1998, Samain2009, Courde-etal:2017} already benefits from operating at 1064\,nm, adopting this wavelength for APOLLO and other stations would enhance the robustness of the diffraction pattern and significantly improve the overall signal-to-noise ratio by mitigating atmospheric attenuation.

Furthermore, the impact of a fixed absolute WFE is strongly wavelength dependent. At 532\,nm, a given WFE constitutes a larger fraction of the wavelength, resulting in more significant phase errors and lower Strehl ratios compared to 1064\,nm. Consequently, hollow CCRs—which typically exhibit WFEs of 30–45\,nm—become very competitive in the near-IR, as the relative phase error is reduced at 1064\,nm, yielding improved diffraction-limited performance.

In addition to these optical advantages, our analysis highlights the significant mass reduction achievable with hollow SiC CCRs. With masses of $\sim$\,0.4–0.5\,kg for a 100\,mm aperture—approximately one-tenth that of solid fused‐silica CCRs (2–2.5\,kg)—hollow designs offer a decisive advantage in payload-mass-constrained missions. Moreover, the high thermal conductivity and moderate thermal expansion of SiC substantially mitigate internal temperature gradients, thereby reducing thermal-lensing effects and preserving optical performance under extreme lunar thermal cycles.

Based on these findings, one may consider deploying two  hollow SiC CCRs, each with a 100\,mm aperture, weighing $\sim$\,0.4--0.5 kg (including athermalized lander fixture), arranged in a dual–CCR configuration co–boresighted with the HGA on the top lander's platform. A modest baseline of $\sim 0.5$\,m between the two CCRs enables differential ranging \cite{Turyshev:CW-LLR:2025} for the purposes of self–calibration and correction of lander–induced thermal expansion errors. In this configuration, each CCR is deliberately offset in its dihedral angles to broaden the overall acceptance and enhance photon return under velocity aberrations and lunar librations. Operating at 1064\,nm further mitigates diffraction losses and minimizes phase errors induced by wavefront distortions. By combining the mass, thermal, and optical advantages of hollow SiC technology with the flexibility of a dual–CCR layout, future LLR experiments can achieve enhanced sensitivity, robust misalignment tolerance, and sustained high–precision ranging in the harsh lunar environment.

In summary, while solid fused‐silica CCRs may offer marginally superior on-axis performance under ideal conditions, the combined advantages of reduced sensitivity to velocity aberration at longer wavelengths, enhanced tolerance to WFEs, and substantially lower mass make hollow CCRs a highly compelling choice for next-generation LLR, particularly under demanding lunar conditions. Our results provide a robust, quantitative foundation for optimizing CCR design—balancing optical throughput, mass efficiency, and thermal stability—to enable high-precision LLR operations \cite{Turyshev:CW-LLR:2025} for geophysical investigations of the Moon and for fundamental tests of gravitation \cite{Williams:2004,Turyshev:2008}.

\section*{Acknowledgments}

The work described here was carried out at the Jet Propulsion Laboratory, California Institute of Technology, Pasadena, California, under a contract with the National Aeronautics and Space Administration.
 \textcopyright 2025. California Institute of Technology. Government sponsorship acknowledged.
 

\begin{thebibliography}{27}
\expandafter\ifx\csname natexlab\endcsname\relax\def\natexlab#1{#1}\fi
\expandafter\ifx\csname bibnamefont\endcsname\relax
  \def\bibnamefont#1{#1}\fi
\expandafter\ifx\csname bibfnamefont\endcsname\relax
  \def\bibfnamefont#1{#1}\fi
\expandafter\ifx\csname citenamefont\endcsname\relax
  \def\citenamefont#1{#1}\fi
\expandafter\ifx\csname url\endcsname\relax
  \def\url#1{\texttt{#1}}\fi
\expandafter\ifx\csname urlprefix\endcsname\relax\def\urlprefix{URL }\fi
\providecommand{\bibinfo}[2]{#2}
\providecommand{\eprint}[2][]{\url{#2}}

\bibitem[{\citenamefont{Dickey et~al.}(1994)\citenamefont{Dickey, Bender,
  Faller, Newhall, Ricklefs, Ries, Shelus, Veillet, Whipple, Wiant
  et~al.}}]{Dickey:1994}
\bibinfo{author}{\bibfnamefont{J.~O.} \bibnamefont{Dickey}},
  \bibinfo{author}{\bibfnamefont{P.~L.} \bibnamefont{Bender}},
  \bibinfo{author}{\bibfnamefont{J.~E.} \bibnamefont{Faller}},
  \bibinfo{author}{\bibfnamefont{X.~X.} \bibnamefont{Newhall}},
  \bibinfo{author}{\bibfnamefont{R.}~\bibnamefont{Ricklefs}},
  \bibinfo{author}{\bibfnamefont{J.~G.} \bibnamefont{Ries}},
  \bibinfo{author}{\bibfnamefont{P.~J.} \bibnamefont{Shelus}},
  \bibinfo{author}{\bibfnamefont{C.}~\bibnamefont{Veillet}},
  \bibinfo{author}{\bibfnamefont{A.~L.} \bibnamefont{Whipple}},
  \bibinfo{author}{\bibfnamefont{J.~R.} \bibnamefont{Wiant}},
  \bibnamefont{et~al.}, \bibinfo{journal}{Science}
  \textbf{\bibinfo{volume}{265}}, \bibinfo{pages}{482} (\bibinfo{year}{1994}).

\bibitem[{\citenamefont{{Murphy} et~al.}(2008)\citenamefont{{Murphy},
  {Adelberger}, {Battat}, {Carey}, {Hoyle}, {LeBlanc}, {Michelsen},
  {Nordtvedt}, {Orin}, {Strasburg} et~al.}}]{Murphy_etal_2008}
\bibinfo{author}{\bibfnamefont{T.~W.} \bibnamefont{{Murphy}},
  \bibfnamefont{Jr.}}, \bibinfo{author}{\bibfnamefont{E.~G.}
  \bibnamefont{{Adelberger}}}, \bibinfo{author}{\bibfnamefont{J.~B.~R.}
  \bibnamefont{{Battat}}}, \bibinfo{author}{\bibfnamefont{L.~N.}
  \bibnamefont{{Carey}}}, \bibinfo{author}{\bibfnamefont{C.~D.}
  \bibnamefont{{Hoyle}}},
  \bibinfo{author}{\bibfnamefont{P.}~\bibnamefont{{LeBlanc}}},
  \bibinfo{author}{\bibfnamefont{E.~L.} \bibnamefont{{Michelsen}}},
  \bibinfo{author}{\bibfnamefont{K.}~\bibnamefont{{Nordtvedt}}},
  \bibinfo{author}{\bibfnamefont{A.~E.} \bibnamefont{{Orin}}},
  \bibinfo{author}{\bibfnamefont{J.~D.} \bibnamefont{{Strasburg}}},
  \bibnamefont{et~al.}, \bibinfo{journal}{PASP} \textbf{\bibinfo{volume}{120}},
  \bibinfo{pages}{20} (\bibinfo{year}{2008}).

\bibitem[{\citenamefont{{Williams} et~al.}(2009)\citenamefont{{Williams},
  {Turyshev}, and {Boggs}}}]{Williams:2009}
\bibinfo{author}{\bibfnamefont{J.~G.} \bibnamefont{{Williams}}},
  \bibinfo{author}{\bibfnamefont{S.~G.} \bibnamefont{{Turyshev}}},
  \bibnamefont{and} \bibinfo{author}{\bibfnamefont{D.~H.}
  \bibnamefont{{Boggs}}}, \bibinfo{journal}{IJMPD}
  \textbf{\bibinfo{volume}{18}}, \bibinfo{pages}{1129} (\bibinfo{year}{2009}).

\bibitem[{\citenamefont{{Murphy} et~al.}(2010)\citenamefont{{Murphy},
  {Adelberger}, {Battat}, {Hoyle}, {McMillan}, {Michelsen}, {Samad}, {Stubbs},
  and {Swanson}}}]{Murphy_etal_2010}
\bibinfo{author}{\bibfnamefont{T.~W.} \bibnamefont{{Murphy}},
  \bibfnamefont{Jr.}}, \bibinfo{author}{\bibfnamefont{E.~G.}
  \bibnamefont{{Adelberger}}}, \bibinfo{author}{\bibfnamefont{J.~B.~R.}
  \bibnamefont{{Battat}}}, \bibinfo{author}{\bibfnamefont{C.~D.}
  \bibnamefont{{Hoyle}}}, \bibinfo{author}{\bibfnamefont{R.~J.}
  \bibnamefont{{McMillan}}}, \bibinfo{author}{\bibfnamefont{E.~L.}
  \bibnamefont{{Michelsen}}}, \bibinfo{author}{\bibfnamefont{R.~L.}
  \bibnamefont{{Samad}}}, \bibinfo{author}{\bibfnamefont{C.~W.}
  \bibnamefont{{Stubbs}}}, \bibnamefont{and}
  \bibinfo{author}{\bibfnamefont{H.~E.} \bibnamefont{{Swanson}}},
  \bibinfo{journal}{Icarus} \textbf{\bibinfo{volume}{208}}, \bibinfo{pages}{31}
  (\bibinfo{year}{2010}).

\bibitem[{\citenamefont{{Currie} et~al.}(2011)\citenamefont{{Currie},
  {Dell'Agnello}, and {Delle Monache}}}]{Currie_etal_2011}
\bibinfo{author}{\bibfnamefont{D.}~\bibnamefont{{Currie}}},
  \bibinfo{author}{\bibfnamefont{S.}~\bibnamefont{{Dell'Agnello}}},
  \bibnamefont{and} \bibinfo{author}{\bibfnamefont{G.}~\bibnamefont{{Delle
  Monache}}}, \bibinfo{journal}{Acta Astronautica}
  \textbf{\bibinfo{volume}{68}}, \bibinfo{pages}{667} (\bibinfo{year}{2011}).

\bibitem[{\citenamefont{Otsubo et~al.}(2011)\citenamefont{Otsubo, Kunimori,
  Noda, Hanada, Araki, and Katayama}}]{Otsubo2011}
\bibinfo{author}{\bibfnamefont{T.}~\bibnamefont{Otsubo}},
  \bibinfo{author}{\bibfnamefont{H.}~\bibnamefont{Kunimori}},
  \bibinfo{author}{\bibfnamefont{H.}~\bibnamefont{Noda}},
  \bibinfo{author}{\bibfnamefont{H.}~\bibnamefont{Hanada}},
  \bibinfo{author}{\bibfnamefont{H.}~\bibnamefont{Araki}}, \bibnamefont{and}
  \bibinfo{author}{\bibfnamefont{M.}~\bibnamefont{Katayama}},
  \bibinfo{journal}{Earth, Planets and Space} \textbf{\bibinfo{volume}{63}},
  \bibinfo{pages}{e13} (\bibinfo{year}{2011}).

\bibitem[{\citenamefont{{Turyshev} et~al.}(2013)\citenamefont{{Turyshev},
  {Williams}, {Folkner}, {Gutt}, {Baran}, {Hein}, {Somawardhana}, {Lipa}, and
  {Wang}}}]{Turyshev-etal:2013}
\bibinfo{author}{\bibfnamefont{S.~G.} \bibnamefont{{Turyshev}}},
  \bibinfo{author}{\bibfnamefont{J.~G.} \bibnamefont{{Williams}}},
  \bibinfo{author}{\bibfnamefont{W.~M.} \bibnamefont{{Folkner}}},
  \bibinfo{author}{\bibfnamefont{G.~M.} \bibnamefont{{Gutt}}},
  \bibinfo{author}{\bibfnamefont{R.~T.} \bibnamefont{{Baran}}},
  \bibinfo{author}{\bibfnamefont{R.~C.} \bibnamefont{{Hein}}},
  \bibinfo{author}{\bibfnamefont{R.~P.} \bibnamefont{{Somawardhana}}},
  \bibinfo{author}{\bibfnamefont{J.~A.} \bibnamefont{{Lipa}}},
  \bibnamefont{and} \bibinfo{author}{\bibfnamefont{S.}~\bibnamefont{{Wang}}},
  \bibinfo{journal}{Experimental Astronomy} \textbf{\bibinfo{volume}{36}},
  \bibinfo{pages}{105} (\bibinfo{year}{2013}).

\bibitem[{\citenamefont{Preston and Merkowitz}(2013)}]{Preston2013}
\bibinfo{author}{\bibfnamefont{A.}~\bibnamefont{Preston}} \bibnamefont{and}
  \bibinfo{author}{\bibfnamefont{S.~M.} \bibnamefont{Merkowitz}},
  \bibinfo{journal}{Applied Optics} \textbf{\bibinfo{volume}{52}},
  \bibinfo{pages}{8676} (\bibinfo{year}{2013}).

\bibitem[{\citenamefont{Degnan}(2023)}]{Degnan:2023}
\bibinfo{author}{\bibfnamefont{J.~J.} \bibnamefont{Degnan}},
  \bibinfo{journal}{Photonics} \textbf{\bibinfo{volume}{10}},
  \bibinfo{pages}{1215} (\bibinfo{year}{2023}).

\bibitem[{\citenamefont{{Turyshev}}(2025)}]{Turyshev:CW-LLR:2025}
\bibinfo{author}{\bibfnamefont{S.~G.} \bibnamefont{{Turyshev}}},
  \bibinfo{journal}{Physical Review Applied} \textbf{\bibinfo{volume}{23}},
  \bibinfo{eid}{064066} (\bibinfo{year}{2025}), \bibinfo{note}{arXiv:2502.02796
  [astro-ph.IM]}.

\bibitem[{\citenamefont{{Williams} et~al.}(2004)\citenamefont{{Williams},
  {Turyshev}, and {Boggs}}}]{Williams:2004}
\bibinfo{author}{\bibfnamefont{J.~G.} \bibnamefont{{Williams}}},
  \bibinfo{author}{\bibfnamefont{S.~G.} \bibnamefont{{Turyshev}}},
  \bibnamefont{and} \bibinfo{author}{\bibfnamefont{D.~H.}
  \bibnamefont{{Boggs}}}, \bibinfo{journal}{Phys. Rev. Lett.}
  \textbf{\bibinfo{volume}{93}}, \bibinfo{eid}{261101} (\bibinfo{year}{2004}).

\bibitem[{\citenamefont{Williams and Boggs}(2020)}]{WilliamsBoggs2020}
\bibinfo{author}{\bibfnamefont{J.~G.} \bibnamefont{Williams}} \bibnamefont{and}
  \bibinfo{author}{\bibfnamefont{D.~H.} \bibnamefont{Boggs}},
  \bibinfo{type}{Interoffice Memorandum} \bibinfo{number}{IOM 335N-20-01},
  \bibinfo{institution}{Jet Propulsion Laboratory, California Institute of
  Technology}, \bibinfo{address}{Pasadena, California} (\bibinfo{year}{2020}).

\bibitem[{\citenamefont{{Williams} et~al.}(2023)\citenamefont{{Williams},
  {Porcelli}, {Dell'Agnello}, {Mauro}, {Muccino}, {Currie}, {Wellnitz}, {Wu},
  {Boggs}, and {Johnson}}}]{Williams-etal:2023}
\bibinfo{author}{\bibfnamefont{J.~G.} \bibnamefont{{Williams}}},
  \bibinfo{author}{\bibfnamefont{L.}~\bibnamefont{{Porcelli}}},
  \bibinfo{author}{\bibfnamefont{S.}~\bibnamefont{{Dell'Agnello}}},
  \bibinfo{author}{\bibfnamefont{L.}~\bibnamefont{{Mauro}}},
  \bibinfo{author}{\bibfnamefont{M.}~\bibnamefont{{Muccino}}},
  \bibinfo{author}{\bibfnamefont{D.~G.} \bibnamefont{{Currie}}},
  \bibinfo{author}{\bibfnamefont{D.}~\bibnamefont{{Wellnitz}}},
  \bibinfo{author}{\bibfnamefont{C.}~\bibnamefont{{Wu}}},
  \bibinfo{author}{\bibfnamefont{D.~H.} \bibnamefont{{Boggs}}},
  \bibnamefont{and} \bibinfo{author}{\bibfnamefont{N.~H.}
  \bibnamefont{{Johnson}}}, \bibinfo{journal}{Planet. Sci. J.}
  \textbf{\bibinfo{volume}{4}}, \bibinfo{eid}{89} (\bibinfo{year}{2023}).

\bibitem[{\citenamefont{Faller and Wampler}(1970)}]{Faller_1970}
\bibinfo{author}{\bibfnamefont{J.~E.} \bibnamefont{Faller}} \bibnamefont{and}
  \bibinfo{author}{\bibfnamefont{E.~J.} \bibnamefont{Wampler}},
  \bibinfo{journal}{Scientific American} \textbf{\bibinfo{volume}{223}},
  \bibinfo{pages}{38} (\bibinfo{year}{1970}).

\bibitem[{\citenamefont{Arnold}(2005)}]{ArnoldApolloAnalysis}
\bibinfo{author}{\bibfnamefont{D.~A.} \bibnamefont{Arnold}},
  \emph{\bibinfo{title}{Cross section of the apollo lunar retroreflector
  arrays}} (\bibinfo{year}{2005}), \bibinfo{note}{{NASA/GSFC Report}},
  \urlprefix\url{https://ilrs.gsfc.nasa.gov/docs/apollo_arrays.pdf}.

\bibitem[{\citenamefont{Malitson}(1965)}]{Malitson1965}
\bibinfo{author}{\bibfnamefont{I.~H.} \bibnamefont{Malitson}},
  \bibinfo{journal}{JOSA} \textbf{\bibinfo{volume}{55}}, \bibinfo{pages}{1205}
  (\bibinfo{year}{1965}).

\bibitem[{\citenamefont{Ciocci et~al.}(2017)\citenamefont{Ciocci, Martini,
  Contessa, Porcelli, Mastrofini, Currie, Delle~Monache, and
  Dell'Agnello}}]{Ciocci2017}
\bibinfo{author}{\bibfnamefont{E.}~\bibnamefont{Ciocci}},
  \bibinfo{author}{\bibfnamefont{M.}~\bibnamefont{Martini}},
  \bibinfo{author}{\bibfnamefont{S.}~\bibnamefont{Contessa}},
  \bibinfo{author}{\bibfnamefont{L.}~\bibnamefont{Porcelli}},
  \bibinfo{author}{\bibfnamefont{M.}~\bibnamefont{Mastrofini}},
  \bibinfo{author}{\bibfnamefont{D.}~\bibnamefont{Currie}},
  \bibinfo{author}{\bibfnamefont{G.}~\bibnamefont{Delle~Monache}},
  \bibnamefont{and}
  \bibinfo{author}{\bibfnamefont{S.}~\bibnamefont{Dell'Agnello}},
  \bibinfo{journal}{Adv. Space Res.} \textbf{\bibinfo{volume}{60}},
  \bibinfo{pages}{1300} (\bibinfo{year}{2017}).

\bibitem[{\citenamefont{Goodrow and Murphy}(2012)}]{Goodrow2012}
\bibinfo{author}{\bibfnamefont{S.~D.} \bibnamefont{Goodrow}} \bibnamefont{and}
  \bibinfo{author}{\bibfnamefont{T.~W.} \bibnamefont{Murphy}},
  \bibinfo{journal}{Applied Optics} \textbf{\bibinfo{volume}{51}},
  \bibinfo{pages}{8793} (\bibinfo{year}{2012}).

\bibitem[{\citenamefont{Born and Wolf}(1999)}]{Born-Wolf:1999}
\bibinfo{author}{\bibfnamefont{M.}~\bibnamefont{Born}} \bibnamefont{and}
  \bibinfo{author}{\bibfnamefont{E.}~\bibnamefont{Wolf}},
  \emph{\bibinfo{title}{Principles of Optics: Electromagnetic Theory of
  Propagation, Interference and Diffraction of Light}}
  (\bibinfo{publisher}{Cambridge University Press}, \bibinfo{year}{1999}),
  \bibinfo{edition}{7th} ed.

\bibitem[{\citenamefont{Goodman}(2017)}]{Goodman:2017}
\bibinfo{author}{\bibfnamefont{J.~W.} \bibnamefont{Goodman}},
  \emph{\bibinfo{title}{Introduction to Fourier Optics}}
  (\bibinfo{publisher}{Roberts and Company Publishers}, \bibinfo{year}{2017}),
  \bibinfo{edition}{4th} ed.

\bibitem[{\citenamefont{{Murphy}}(2013)}]{Murphy:2013}
\bibinfo{author}{\bibfnamefont{T.~W.} \bibnamefont{{Murphy}},
  \bibfnamefont{Jr.}}, \bibinfo{journal}{Rep. Progr. Phys.}
  \textbf{\bibinfo{volume}{76}}, \bibinfo{eid}{076901} (\bibinfo{year}{2013}).

\bibitem[{\citenamefont{Murphy and Goodrow}(2013)}]{Murphy2013Polarization}
\bibinfo{author}{\bibfnamefont{T.~W.} \bibnamefont{Murphy}} \bibnamefont{and}
  \bibinfo{author}{\bibfnamefont{S.~D.} \bibnamefont{Goodrow}},
  \bibinfo{journal}{Applied Optics} \textbf{\bibinfo{volume}{52}},
  \bibinfo{pages}{117} (\bibinfo{year}{2013}).

\bibitem[{\citenamefont{{Battat} et~al.}(2009)\citenamefont{{Battat}, {Murphy},
  {Adelberger}, {Gillespie}, {Hoyle}, {McMillan}, {Michelsen}, {Nordtvedt},
  {Orin}, {Stubbs} et~al.}}]{Battat_etal_2009}
\bibinfo{author}{\bibfnamefont{J.~B.~R.} \bibnamefont{{Battat}}},
  \bibinfo{author}{\bibfnamefont{T.~W.} \bibnamefont{{Murphy}}},
  \bibinfo{author}{\bibfnamefont{E.~G.} \bibnamefont{{Adelberger}}},
  \bibinfo{author}{\bibfnamefont{B.}~\bibnamefont{{Gillespie}}},
  \bibinfo{author}{\bibfnamefont{C.~D.} \bibnamefont{{Hoyle}}},
  \bibinfo{author}{\bibfnamefont{R.~J.} \bibnamefont{{McMillan}}},
  \bibinfo{author}{\bibfnamefont{E.~L.} \bibnamefont{{Michelsen}}},
  \bibinfo{author}{\bibfnamefont{K.}~\bibnamefont{{Nordtvedt}}},
  \bibinfo{author}{\bibfnamefont{A.~E.} \bibnamefont{{Orin}}},
  \bibinfo{author}{\bibfnamefont{C.~W.} \bibnamefont{{Stubbs}}},
  \bibnamefont{et~al.}, \bibinfo{journal}{PASP} \textbf{\bibinfo{volume}{121}},
  \bibinfo{pages}{29} (\bibinfo{year}{2009}).

\bibitem[{\citenamefont{Samain et~al.}(1998)\citenamefont{Samain, Veillet,
  Fridelance, and {et al.}}}]{Samain1998}
\bibinfo{author}{\bibfnamefont{E.}~\bibnamefont{Samain}},
  \bibinfo{author}{\bibfnamefont{P.}~\bibnamefont{Veillet}},
  \bibinfo{author}{\bibfnamefont{C.}~\bibnamefont{Fridelance}},
  \bibnamefont{and} \bibinfo{author}{\bibnamefont{{et al.}}},
  \bibinfo{journal}{Astron. Astrophys.} \textbf{\bibinfo{volume}{336}},
  \bibinfo{pages}{L17} (\bibinfo{year}{1998}).

\bibitem[{\citenamefont{Samain et~al.}(2009)\citenamefont{Samain, Guillemot,
  Courde, and Torre}}]{Samain2009}
\bibinfo{author}{\bibfnamefont{E.}~\bibnamefont{Samain}},
  \bibinfo{author}{\bibfnamefont{P.}~\bibnamefont{Guillemot}},
  \bibinfo{author}{\bibfnamefont{C.}~\bibnamefont{Courde}}, \bibnamefont{and}
  \bibinfo{author}{\bibfnamefont{L.}~\bibnamefont{Torre}}, in
  \emph{\bibinfo{booktitle}{16th International Workshop on Laser Ranging}}
  (\bibinfo{year}{2009}),
  \urlprefix\url{https://ilrs.gsfc.nasa.gov/lw16/docs/papers/new_4_Samain_p.pdf}.

\bibitem[{\citenamefont{Courde et~al.}(2017)\citenamefont{Courde, Torre,
  Samain, Martinot-Lagarde, Aimar, Albanese, Exertier, Fienga, Mariey, Metris
  et~al.}}]{Courde-etal:2017}
\bibinfo{author}{\bibfnamefont{C.}~\bibnamefont{Courde}},
  \bibinfo{author}{\bibfnamefont{J.~M.} \bibnamefont{Torre}},
  \bibinfo{author}{\bibfnamefont{E.}~\bibnamefont{Samain}},
  \bibinfo{author}{\bibfnamefont{G.}~\bibnamefont{Martinot-Lagarde}},
  \bibinfo{author}{\bibfnamefont{M.}~\bibnamefont{Aimar}},
  \bibinfo{author}{\bibfnamefont{D.}~\bibnamefont{Albanese}},
  \bibinfo{author}{\bibfnamefont{P.}~\bibnamefont{Exertier}},
  \bibinfo{author}{\bibfnamefont{A.}~\bibnamefont{Fienga}},
  \bibinfo{author}{\bibfnamefont{H.}~\bibnamefont{Mariey}},
  \bibinfo{author}{\bibfnamefont{G.}~\bibnamefont{Metris}},
  \bibnamefont{et~al.}, \bibinfo{journal}{Astron. Astrophys.}
  \textbf{\bibinfo{volume}{602}}, \bibinfo{pages}{A90} (\bibinfo{year}{2017}).

\bibitem[{\citenamefont{{Turyshev}}(2008)}]{Turyshev:2008}
\bibinfo{author}{\bibfnamefont{S.~G.} \bibnamefont{{Turyshev}}},
  \bibinfo{journal}{Ann. Rev. Nucl. Part. Sci.} \textbf{\bibinfo{volume}{58}},
  \bibinfo{pages}{207} (\bibinfo{year}{2008}).

\end{thebibliography}

\end{document}